\documentclass[aps,reprint,superscriptaddress,floatfix]{revtex4-1}

\expandafter\let\csname equation*\endcsname\relax 
\expandafter\let\csname endequation*\endcsname\relax
\usepackage{amsmath}
\makeatletter
\@namedef{ver@amsmath.sty}{}
\makeatother
\usepackage{amstext}
\usepackage[version=3]{mhchem}
\usepackage{graphicx}
\usepackage{nicefrac}
\usepackage{siunitx}
\usepackage{url}

\begin{document}
	
\title{Stability and accuracy control of k.p parameters}

\author{Carlos M. O. Bastos}
\address{S\~ao Carlos Institute of Physics, University of S\~ao Paulo, PO Box 369, 13560-970, S\~ao Carlos, SP, Brazil.}

\author{Fernando P. Sabino}
\address{S\~ao Carlos Institute of Physics, University of S\~ao Paulo, PO Box 369, 13560-970, S\~ao Carlos, SP, Brazil.}

\author{Paulo E. Faria Junior}
\address{S\~ao Carlos Institute of Physics, University of S\~ao Paulo, PO Box 369, 13560-970, S\~ao Carlos, SP, Brazil.}
\address{Department of Physics, State University of New York at Buffalo, 14260, Buffalo, New York, USA.}

\author{Tiago Campos}
\address{S\~ao Carlos Institute of Physics, University of S\~ao Paulo, PO Box 369, 13560-970, S\~ao Carlos, SP, Brazil.}

\author{Juarez L. F. Da Silva}
\address{S\~ao Carlos Institute of Chemistry, University of S\~ao Paulo, PO Box 780, 13560-970, S\~ao Carlos, SP, Brazil.}

\author{Guilherme M. Sipahi}
\address{S\~ao Carlos Institute of Physics, University of S\~ao Paulo, PO Box 369, 13560-970, S\~ao Carlos, SP, Brazil.}
\address{Department of Physics, State University of New York at Buffalo, 14260, Buffalo, New York, USA.}

\begin{abstract}
The $\mathbf{k{\cdot}p}$ method is a successful approach to obtain band structure, 
optical and transport properties of semiconductors, and it depends on external parameters 
that are obtained either from experiments, tight binding or \textit{ab initio} calculations. 
Despite the widespread use of the $\mathbf{k{\cdot}p}$ method, a systematic analysis of the stability 
and the accuracy of its parameters is not usual in the literature.
In this work, we report a theoretical framework to determine the $\mathbf{k{\cdot}p}$ parameters
from state-of-the-art hybrid density functional theory including 
spin-orbit coupling, providing a calculation where the gap and spin-orbit 
energy splitting are 
in agreement with the experimental values.
The accuracy of the set of parameters is enhanced by fitting over several directions at once, 
minimizing the overall deviation from the original data. This strategy allows 
us to systematically
evaluate the stability, preserving the accuracy of the parameters, providing a tool to determine optimal
parameters for specific ranges around the $\Gamma$-point.
To prove our concept, we investigate the zinc blende GaAs that shows results
in excellent agreement with the most reliable data in the literature.
\end{abstract}


\maketitle

\section{Introduction}
\label{sec:introd}

A deep knowledge of the band structure (electronic states) of semicondutors 
is one of the first steps towards the understanding of a wide range of physical 
systems and phenomena, such as topological 
insulators~\cite{Bernevig.Science.314.5806,Baum.PhysRevB.89.245136,Miao2012}, 
Majorana 
fermions~\cite{Alicea2012,Mourik.Science.336.1003,Stanescu.JPCM.23.233201,Reuther.PhysRevX.3.031011}
 and polytypic nanowhiskers~\cite{FariaJunior2012,FariaJunior2014} or 
technologies such as 
spintronics~\cite{Jungwirth.RevModPhys.86.855,Lazic.PhysRevB.90.085429}. 
The band structure of a given material of interest can be obtained using experimental 
information~\cite{Kane1957,Walukiewicz.JPhysD:Appl.39.R83} or based on 
theoretical calculations using different level of approximations developed 
along the years \cite{Jones.RevModPhys.87.897}, i.e., the effects of particular 
interactions can be studied in details. For example, the role of the spin-orbit 
coupling (SOC) can be studied in detail using different approximations, which 
is crucial as SOC plays a critical role in the systems mentioned above.  

The theoretical approaches to calculate the band structure for a given material 
can be separated in two lines, namely, $(i)$ first-principles methods based on 
density functional theory (DFT)~\cite{Hohenberg1964,Kohn1965} or 
quantum-chemistry methods such as the Hartree-Fock; $(ii)$ phenomenological 
approaches such as the  
$\mathbf{k{\cdot}p}$~\cite{Kane1966,Sipahi.PhysRevB.57.9168} or tight-binding 
\cite{Slater.PhysRev.94.1498,harrison1989,Goringe.RepProgPhys.60.1447} methods. 
For crystalline materials, both first-principles and phenomenological 
approaches can be applied and their results can be compared with experiments, 
and hence, their accuracy can be established. However, the use of 
first-principles methods for modelling confined systems such as quantum-dots, 
nanowires, etc, requires supercells with thousand or even million atoms, which 
are forbidden because of its computational cost. In contrast, the 
$\mathbf{k{\cdot}p}$ method has a lower computational cost because the 
interactions between the particles are described by an effective 
potential set up by a set of parameters. The determination of such parameters 
is of seminal importance. 

The $\mathbf{k{\cdot}p}$ Hamiltonian is constructed using the framework of 
perturbation theory~\cite{Enderlein1997,Willatzen2009} and group theory 
analysis to reduce the number of matrix elements that are replaced by effective 
parameters. The number of parameters depend on the number of selected bands 
and on the symmetry of the described crystal. In zinc blende crystals, there is 
a relation allowing to calculate the effective mass parameters using the 
effective masses themselves~\cite{Enderlein1998}, but for wurtzite crystal 
symmetry this is no longer true~\cite{Chuang&Chang.PhysRevB.54.2491}. 

The effective masses can be determined experimentally using, for example, 
cyclotron resonance~\cite{Mears1971,Herlach1974}, Hall effect 
\cite{Becker.JApplPhys.32.2094} or optical measurements 
\cite{Spitzer.PhysRev.106.882,Cardona.PhysRev.121.752}, or theoretically, 
fitting a parabolic dispersion very close to $\Gamma$-point of \textit{ab 
initio} band structure calculations~\cite{Dugdale2000,Ramos2001}. These 
procedures are only able to produce the effective mass parameters, leaving to 
other techniques the task of setting the values for the interband coupling 
parameters, such as the well known Kane parameter, $P$. This parameter is 
usually extracted from the effective g factor~\cite{Vurgaftman2001}. 

Parameters for most of the standard compounds may be found on the 
literature~\cite{Vurgaftman2001,Boujdaria2001,Vurgaftman2003,Shokhovets2007}. 
For example, Ref.~\cite{Vurgaftman2001} presents a compilation of parameters 
for almost all binary, ternary and quaternary zinc blende compounds and also 
for wurtzite III-nitrides, however, those parameters were obtained by mixing 
experimental and theoretical data, i.e., no systematic procedure was employed. 
The reference provides parameters for the usual $6{\times}6$ 
(Luttinger-Kohn~\cite{Luttinger1955}) and $8{\times}8$ (Kane~\cite{Kane1966} 
and Rashba-Sheka-Pikus~\cite{Sirenko.PhysRevB.53.1997}) band models. As we go 
further into the $\mathbf{k{\cdot}p}$ models, there exist only few reliable 
sources of parameters for models with a higher number of bands, e.g., 
for $14{\times}14$ \cite{Jancu2005,Winkler2003}, 
$20{\times}20$~\cite{Radhia2002}, 
$24{\times}24$~\cite{Radhia2003}, $34{\times}34$~\cite{Saidi2008} and $40{\times}40$~\cite{Saidi2010} 
bands. 

In this paper, we developed a new framework to determine the 
$\mathbf{k{\cdot}p}$ parameters from preexistent band structures that works 
with any crystal symmetry. Fitting a set of functions derived from the secular 
equation of the $\mathbf{k{\cdot}p}$ Hamiltonian to a preexistent band 
structure we were able to extract all the $\mathbf{k{\cdot}p}$ parameters at 
once, including the interband coupling parameters. Furthermore, we performed 
the fitting using several different directions of the first Brillouin zone 
(FBZ), thus finding all the parameters in a consistent way. As a proof of 
concept we use a zinc blende GaAs band structure obtained by the hybrid DFT 
calculation with the Heyd-Scuseria-Ernzerhof functional (HSE). Since GaAs is 
the most studied material it will be easy to compare our results with the 
reported values in the literature. Furthermore, using our method we are able to 
predict the best set of parameters for a specific region of the FBZ. We show 
that, in the GaAs case, our parameters are in good agreement with the 
literature. We also address the accuracy of the Kane model by defining 
an strategy to evaluate its limits of validity. In conclusion, our method is 
neither limited to the crystal phase nor the Hamiltonian and opens up the 
possibility to study novel semiconductor systems.

The paper is organized as follows: In section \ref{sec:kp} we present the 
the 8$\times$8 $\mathbf{k{\cdot}p}$ Hamiltonian. The process to obtain 
DFT-HSE band structure for GaAs is shown in section \ref{sec:dftGaAs}. 
In section \ref{sec:fittingmethod} we show the developed method for a general 
Hamiltonian specify the expressions for the 8$\times$8 Hamiltonian. The 
application of the method to zinc blende GaAs band structure is presented in 
\ref{sec:kpGaAs}. We proceed to the analysis of optimal parameters in section 
\ref{sec:optimizationGaAs}, comparing our results with the literature in 
section \ref{sec:comparison}. Finally, our conclusions are shown in 
section~\ref{sec:conclusion}. 
 

\section{The $\mathbf{k{\cdot}p}$ method} \label{sec:kp}

In this paper, we employed the $8{\times}8$ $\mathbf{k{\cdot}p}$ Hamiltonian proposed by Kane~\cite{Kane1966}, 
that extends the $6{\times}6$ Hamiltonian proposed by 
Luttinger-Kohn~\cite{Luttinger1955}, in which the first-order contribution of the $\mathbf{k}$-dependent spin-orbit term  
and also the second order contribution of $\mathbf{k{\cdot}p}$ between conduction (CB) and valence (VB) bands are neglected.
Further details on the $\mathbf{k{\cdot}p}$ Hamiltonian are discussed in the Supplemental Material.
The $8\times8$ Kane Hamiltonian shows as

\begin{widetext}
\begin{equation}
 \begin{pmatrix}
  Q & S & R & 0 & i\frac{S}{\sqrt{2}} & -i\sqrt{2}R & -iP_-& 0\\
  S^{\dagger} & T & 0 & R & i\frac{(T-Q)}{\sqrt{2}} & i\sqrt{\frac{3}{2}}S &
  \sqrt{\frac{2}{3}}P_z & -\frac{1}{\sqrt{3}}P_-\\
  R^{\dagger} & 0 & T & -S & -i\sqrt{\frac{3}{2}}S^{\dagger} & i\frac{(T-Q)}{\sqrt{2}} &
  -\frac{i}{\sqrt{3}}P_+ & -i\sqrt{\frac{2}{3}}P_z \\
  0 & R^{\dagger} & -S^{\dagger} & Q & -i\sqrt{2}R^{\dagger} & 
  -i\frac{S^{\dagger}}{\sqrt{2}}& 0 & -P_+\\
  -i\frac{S^{\dagger}}{\sqrt{2}} & -i\frac{(T-Q)^{\dagger}}{\sqrt{2}} &
  i\sqrt{\frac{3}{2}}S & i\sqrt{2}R & \frac{Q+T}{2}+\Delta_{so} & 0 & 
  -\frac{i}{\sqrt{3}}P_z &-i\sqrt{\frac{2}{3}}P_- \\
  i\sqrt{2}R^{\dagger} & -i\sqrt{\frac{3}{2}}S^{\dagger} &
  -i\frac{(T-Q)^{\dagger}}{\sqrt{2}} & i\frac{S}{\sqrt{2}} & 0 & 
  \frac{Q+T}{2}+\Delta_{so} & \sqrt{\frac{2}{3}}P_+ &  -\frac{1}{\sqrt{3}}P_z\\
  -iP_-& \sqrt{\frac{2}{3}}P_z  &  \frac{i}{\sqrt{3}}P_-& 0&  
  \frac{i}{\sqrt{3}}P_z & \sqrt{\frac{2}{3}}P_- & E_c & 0\\
  0 &  - \frac{1}{\sqrt{3}}P_+ & i\sqrt{\frac{2}{3}}P_z  & -P_- & 
  i\sqrt{\frac{2}{3}}P_+ &  -\frac{1}{\sqrt{3}}P_z & 0 & E_c \\ 
 \end{pmatrix}\ ,
\label{eq:kpZB8}
\end{equation}
where the terms are given by
\begin{equation}
 \begin{aligned}[l]
  Q&=-\frac{\hbar^2}{2 m_0}\left[(\tilde{\gamma}_{1}+\tilde{\gamma}_{2})(k_{x}^{2}+k_{y}^{2})-
  (\tilde{\gamma}_{1}-2\tilde{\gamma}_{2})\, k_{z}^{2}\right] &
  R&=-\frac{\hbar^2}{2 m_0}\sqrt{3}\left[
  \tilde{\gamma}_{2} (k_{x}^{2}-k_{y}^{2}) + 2 i \tilde{\gamma}_{3} k_{x} k_{y}
  \right]\\
  E_c&=E_g + \frac{\hbar^2}{2 m_0} \tilde{\text{e}}\, k^2 &
  P_z&=\text{P} \, k_z \\
  T&=-\frac{\hbar^2}{2 m_0}\left[(\tilde{\gamma}_{1}-\tilde{\gamma}_{2})(k_{x}^{2}+k_{y}^{2})+
  (\tilde{\gamma}_{1}+2\tilde{\gamma}_{2})\, k_{z}^{2} \right]&
  S&=i\frac{\hbar^2}{2 m_0}\left[2 \sqrt{3}\tilde{\gamma}_{3}k_{z}(k_{x}-ik_{y})\right]\\
  P_\pm&=\frac{1}{\sqrt{2}}\text{P}\left(k_x\pm ik_y\right) &
  k^2&=k_x^2+k_y^2+k_z^2\\
 \end{aligned}\ 
\end{equation}
\end{widetext}
with the following parameters: 
\begin{itemize}
\item $\tilde{\gamma}_1$, $\tilde{\gamma}_2$, $\tilde{\gamma}_3$, $\tilde{e}$: 
second order effective mass parameters of VB and CB. These parameters are 
adimensional~\footnote{The tilde is used, as in $\tilde{\gamma}_1$, to refer to the Kane model parameters 
in opposition to the Luttinger parameters, defined in Ref.~\cite{Luttinger1955}, that are noted without it, as in $\gamma_1$.}
.
\item $\text{P}$: first order interaction term between states in the conduction and the 
valence bands. An energy equivalent, $E_P = \nicefrac{2 m_0 P^2}{\hbar^2}$, may be used to analyze 
the effects of this parameter. 
\item $\Delta_{so}$: first order SOC interaction term (energy difference between 
HH/LH and SO bands at $\Gamma$-point).
\item $E_g$: energy band gap between the CB and HH/LH bands at the 
$\Gamma$-point. 
\end{itemize}

The Kane Hamiltonian basis set is composed by the topmost six states of VB and 
the first two states of CB, in the following order:
$\left|\text{HH}\Uparrow\right\rangle$, 
$\left|\text{LH}\Uparrow\right\rangle$, 
$\left|\text{LH}\Downarrow\right\rangle$, 
$\left|\text{HH}\Downarrow\right\rangle$, 
$\left|\text{SO}\Uparrow\right\rangle$, 
$\left|\text{SO}\Downarrow\right\rangle$,
$\left|\text{CB}\Uparrow\right\rangle$,
$\left|\text{CB}\Downarrow\right\rangle$.
HH, LH, SO and CB stand for the heavy hole, light hole, split-off hole and conduction 
band states, respectively. 
$\Uparrow$ and $\Downarrow$ are use to distinguish the total angular momentum projections. 
The states description using the original atomic orbital basis is given by:
\begin{equation}
\small
\begin{aligned}[l]
  \left|\text{HH}\Uparrow\right\rangle   &
  =\frac{1}{\sqrt{2}}\left|\left(X+iY\right)\uparrow\right\rangle \\
  \left|\text{HH}\Downarrow\right\rangle  &
  =\frac{i}{\sqrt{2}}\left|\left(X-iY\right)\downarrow\right\rangle \\
  \left|\text{LH}\Uparrow\right\rangle  &
  =\frac{i}{\sqrt{6}}\left|\left(X+iY\right)\downarrow-2Z\uparrow\right\rangle \\
  \left|\text{LH}\Downarrow\right\rangle  &
  =\frac{1}{\sqrt{6}}\left|\left(X-iY\right)\uparrow+2Z\downarrow\right\rangle \\
  \left|\text{SO}\Uparrow\right\rangle  &
  =\frac{1}{\sqrt{3}}\left|\left(X+iY\right)\downarrow+Z\uparrow\right\rangle \\
  \left|\text{SO}\Downarrow\right\rangle  &
  =\frac{i}{\sqrt{3}}\left|-\left(X-iY\right)\uparrow+Z\downarrow\right\rangle \\
  \left|\text{CB}\Uparrow\right\rangle  & =
  \left|S\uparrow\right\rangle \\
   \left|\text{CB}\Downarrow\right\rangle  & =\left|S\downarrow\right\rangle \\
 \end{aligned}
\label{eq:basekp}
\end{equation}
where 
$\left|X\right\rangle$, $\left|Y\right\rangle$ and $\left|Z\right\rangle$
are the p-type like states ($p_x$, $p_y$ and $p_z$) and 
$\left|S\right\rangle$  the s-like ones and $\uparrow$ and 
$\downarrow$ represent their spins.


\section{G\MakeLowercase{a}A\MakeLowercase{s} hybrid DFT-HSE band structure} \label{sec:dftGaAs}

The {\it ab initio} \ce{GaAs} band structure 
was obtained through the use of hybrid DFT calculations within the HSE~\cite{Heyd2003} 
exchange-correlation (XC) functional where the energy is given by the following 
equation, 
\begin{equation}
E_{\text{xc}}^{\text{HSE}} = \alpha E_{\text{x}}^{\text{SR}}(\mu) + 
(1-\alpha)E_{\text{x}}^{\text{PBE,SR}} + E_{\text{x}}^{\text{PBE,LR}} + 
E_{\text{c}}^{\text{PBE}}\, . 
\label{eq:energy_HSE}
\end{equation}

In the HSE formulation, the exchange energy is partitioned into two terms, namely, 
short range (SR) and long range (LR) terms. A screening nonlocal Fock operator 
is employed to obtain the SR term, in which the $\mu$ parameter ($\mu= \SI{0.206}{\AA}^{-1}$) 
determines the intensity of the screening \cite{Heyd2004}, while the LR term is 
described by the semilocal Perdew-Burke-Ernzerhof (PBE) \cite{Perdew1996} functional. 
The parameter $\alpha$ defines the percentage of the nonlocal SR exchange term, 
and it is \SI{0.25}{} in the HSE06 functional. However, this particular value was 
obtained for typical molecules, based on the analysis of the adiabatic connection 
formula and the lowest order of G\"orling-Levy perturbation theory \cite{Perdew1996}, 
and consequently does not yield a correct band gap (although better than PBE) for 
most of the materials \cite{Perdew1996,Heyd2004,Moses2011}. Therefore, the $\alpha$ 
parameter can be assumed as a fitting parameter, which can be adjusted to reproduce
 particular bulk properties, e.g., energy band gap, lattice parameter, etc. We 
have fitted $\alpha$ to yield the fundamental experimental \ce{GaAs} band 
gap~\cite{Vurgaftman2001,Willatzen2009} (i.e., \SI{1.519}{\eV}).

To solve the Kohn-Sham equation, we employed the projected augmented wave (PAW) 
method \cite{Blochl1994,Kresse1999}, as implemented in Vienna \textit{Ab-initio} 
Simulation Package (VASP) \cite{Kresse1993,Kresse1996}, and the PAW projectors 
provided within VASP to describe the following valence states, $4s^{2}4p^{3}$ for 
\ce{As} and $3d^{10}4s^{2}4p^{1}$ for \ce{Ga}. To describe the valence electronic 
states, we employed the scalar relativistic approximation, in which the SOC effects 
for the valence states were taken into account by perturbation theory employing 
\num{38} empty states. For the total energy and band structure calculations, we 
employed a cutoff energy of \SI{455}{\eV}, while a cutoff energy of \SI{607}{\eV} 
was used to obtain the equilibrium volume by the minimization of the stress tensor. 
For the Brillouin zone integration, we employed a $\mathbf{k}$-point mesh of $8{\times}8{\times}8$, 
which yields accurate results. 
 
\ce{GaAs} crystallize in the well known zinc blende structure with space group 
$T_d$ and one formula unit per primitive unit cell, in which every \ce{Ga} atom 
is surrounded by four \ce{As} atoms (tetrahedral symmetry), and vice-versa. 
Using the fact that the band gap of a semiconductor increases almost linearly by increasing 
the percentage of the $\alpha$ (nonlocal Fock term) parameter \cite{Moses2011} 
we set $\alpha = \SI{0.317}{}$, finding a \ce{GaAs} band gap of \SI{1.521}{\eV}, 
which deviates by \SI{14}{\percent} from the experimental result. We obtained an 
equilibrium lattice parameter of \SI{5.652}{\AA}, which deviates by \SI{1}{\percent} 
compared with the experimental results. However, we would like to point out that 
using $\alpha = \SI{0}{}$ yields $a_0 = \SI{5.733}{\AA}$. Furthermore, total calculations 
at the same lattice constant without the SOC for the valence states increase the 
band gap to $\SI{1.627}{\eV}$.

\begin{figure}[!htbp]
\centering
\includegraphics[width=0.48\textwidth]{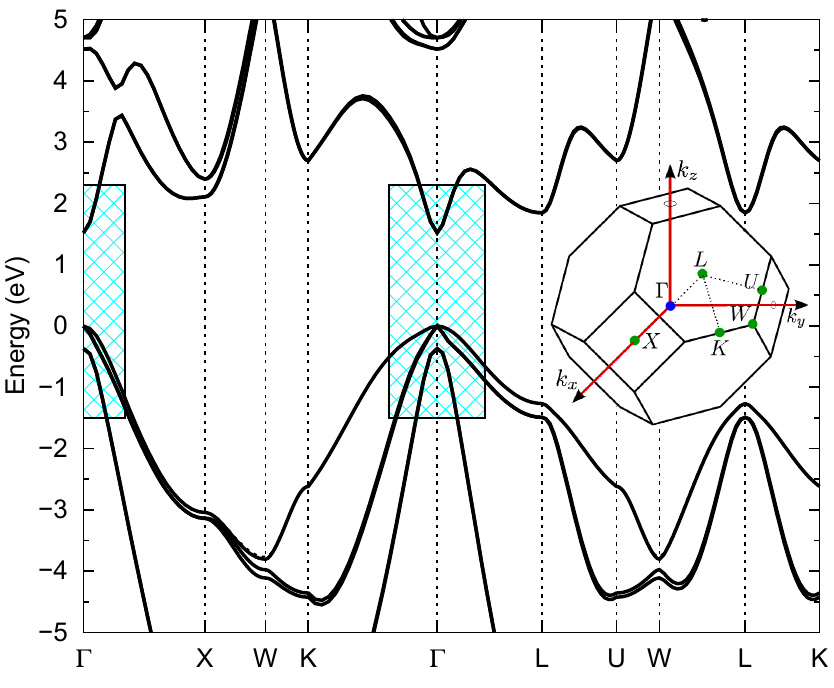}
\caption{Band structure for \ce{GaAs} zinc blende with the respective
FBZ and the high symmetry points. The highlighted regions indicate the approximate 
region where Kane Hamiltonian is valid.}
\label{fig:band_structure}
\end{figure}

Fig.~\ref{fig:band_structure} presents the band structure with SOC for the 
valence states on the usual high-symmetry Brillouin lines. Both the VBM and the 
CBM are located at the $\Gamma\textrm{-point}$, as expected 
\cite{Vurgaftman2001}. In the absence of SOC (not shown here), the 
highest valence band is composed by a sixfold degenerate state (each band being 
twofold degenerate in spin) at $\Gamma$-point. However, when the SOC is 
considered this degeneracy is broken in a twofold degenerate, split-off band, 
and a fourfold degenerate band that still remains the highest valence band. The 
energy difference between these states are $\Delta_{so} = \SI{0.369}{\eV}$ 
which is \SI{8}{\percent} bigger than the experimental value, \SI{0.341}{\eV} 
\cite{Vurgaftman2001}. Even thought the band gap was fitted to yield the 
experimental result, the $\Delta_{so}$ parameter was not fitted in our 
calculations, which explains this difference. For $\mathbf{k}$-points 
other than $\Gamma$, SOC still breaks the degeneracy of the highest valence 
band, creating the heavy and light holes bands. A closer look will show that in 
less symmetric points, e. g. along the $\Gamma-\text{K}$ line, the degeneracy 
is further broken, creating bands with no degeneracy. The present results are 
consistent with the literature, including SOC or not. 

To ensure high quality data for the fitting along the desired $\Gamma-\text{X}$, 
$\Gamma-\text{L}$ and $\Gamma-\text{K}$ lines, we calculated a large number of $\mathbf{k}$-points 
along each line. Because of the perturbation theory used in the $\mathbf{k{\cdot}p}$ 
method, we expect to fit the parameters only at a defined region around the $\Gamma$-point. 
In order to reduce the high computational cost of the hybrid DFT-HSE+SOC approach, 
we restricted $\mathbf{k}$-points to \num{100} samples up to \SI{50}{\percent} 
of the FBZ. 
   
\section{The fitting method} \label{sec:fittingmethod}

Along the years, $\mathbf{k{\cdot}p}$ parameters are usually being derived from
effective masses using experimental
data~\cite{Vrehen1968,Molenkamp1988,Neumann1988,Binggeli1991,Shokhovets2007}
or from theoretical band structure calculations
\cite{Lawaetz1971,Kim1997a,Yeo1998,Ramos2001,Rezaei2006,Kim2010,Cheiwchanchamnangij2011}. 
Although this procedure is relatively simple, it is not always possible to find
analytical solutions relating $\mathbf{k{\cdot}p}$ parameters to the effective masses.
Alternatively, the determination of the parameters may rely on the fitting of previously 
calculated band structures\cite{Suzuki1995,Yeo1998,Pugh1999a,Ren1999b,Dugdale2000, %
Fritsch2003,Rinke2008,Punya2012}.
However, details about the fitting approach are not usually described by the 
 authors.

 Although we consider a fitting that is based on the resolution of the secular 
 equation, as in previous works \cite{ Kim1997a, Rinke2008,Cardona1966}, 
 we used it in a different way.
 The secular equation is used to reduce the complexity of the fitting. 
 Using the property that the eigenvalues are roots of the secular equation and, 
 by consequence, assuming that we can collect expressions for any of the 
 coefficients  that must also be zero, we extract a new set of $n$ equations 
 ($n$  being the order of the original matrix) that are used in our fitting. 
 With this procedure, we make explicit the couplings among the  different 
 bands,  simplifying the expressions to be solved. The functions determined for 
 each direction are used together to provide the  fitting that minimizes the 
 euclidean distance of the full set of equations to  the  previously calculated 
 data at once.  Because we determine all the  distances in a single step, we 
 guarantee that no  direction is assigned more  importance than any other. In 
 fact, the addition of  other directions to the fitting provides a way to 
 increase the accuracy.

The general form of any $\mathbf{k{\cdot}p}$ matrix with
$n$ energy bands is given by
\begin{equation}
  \left(\begin{array}{ccccc}
  \alpha_{11}(\mathbf{k},\{p\}) & \ldots & \alpha_{1i}(\mathbf{k},\{p\}) &
  \ldots & \alpha_{1n}(\mathbf{k},\{p\})\\
  \vdots & \ddots & \vdots & \ddots & \vdots\\
  \alpha_{1i}^\dagger(\mathbf{k},\{p\}) & \ldots & \alpha_{ii}(\mathbf{k},\{p\}) 
  & \ldots & \alpha_{in}(\mathbf{k},\{p\})\\
  \vdots & \ddots & \vdots & \ddots & \vdots\\
  \alpha_{1n}^\dagger(\mathbf{k},\{p\}) & \ldots & \alpha_{in}^\dagger(\mathbf{k},\{p\})
  & \ldots & \alpha_{nn}(\mathbf{k},\{p\})
  \end{array}\right) \, ,
  \label{eq:Generalhamiltonian}
 \end{equation}
where the matrix elements, $\alpha_{ij}(\mathbf{k},\{p\})$ are functions that 
represent each matrix element with $\mathbf{k}$ being the wave 
vector and $\{p\}$, the set of $\mathbf{k{\cdot}p}$ parameters to be determined.

The secular equation of the Hamiltonian (\ref{eq:Generalhamiltonian}) 
may be written as a general polynomial for the eigenvalues $\epsilon$
\begin{equation}
 \begin{split}
  & c_{n-1}\left(\alpha_{11}(\mathbf{k},\{p\}),...,\alpha_{nn}(\mathbf{k},
  \{p\})\right)\epsilon^{n-1}(\mathbf{k})\\
  & + ... + c_{1}\left(\alpha_{11}(\mathbf{k},\{p\}),...,\alpha_{nn}
  (\mathbf{k},\{p\})\right)\epsilon(\mathbf{k}) \\
  & + c_{0}\left(\alpha_{11}(\mathbf{k},\{p\}),...,\alpha_{nn}(\mathbf{k},\{p\})
  \right) = -\epsilon^{n}(\mathbf{k}) , \, 
 \end{split}
\end{equation}
where $c_i$ are the polynomial coefficients, functions of the matrix elements 
$\alpha_{ij}(\mathbf{k},\{p\})$. Since these coefficients are functions of 
$\mathbf{k}$ and $\{p\}$, we can denote them as $c_i(\mathbf{k},\{p\})$, 
rewriting the above equation as 
\begin{equation}
 \sum^{n-1}_{i=0} c_{i}(\mathbf{k},\{p\}) \epsilon^i(\mathbf{k}) = -\epsilon^n(\mathbf{k})  \, .
 \label{eq:secpoly}
\end{equation}

The analytical forms of these coefficients are used as the fitting functions on 
our approach, and will be identified as \textit{analytical functions}, 
denoted by the super-index $A$:
\begin{equation}
 \begin{cases}
  c^A_0(\mathbf{k},\{p\})\\
  c^A_1(\mathbf{k},\{p\})\\
  \vdots\\
  c^A_{n-1}(\mathbf{k},\{p\}) \, .
 \end{cases}
\end{equation}

The next step is to find a similar relation for the eigenvalues 
obtained from the preexistent band structures, from now on 
called \textit{reference band structure}.
Assuming that the eigenvalues satisfy the secular equation, we can 
write a system of equations to determine the polynomial 
coefficients as a function of the wave vector $\mathbf{k}$:
\begin{equation}
\small
 \left(
 \begin{array}{cccccc}
  1 & \epsilon_{1}(\mathbf{k}) & \ldots & \epsilon_{1}^{i}(\mathbf{k}) & \ldots
  & \epsilon_{1}^{n-1}(\mathbf{k})\\
  \vdots & \vdots & \ddots & \vdots & \ddots & \vdots\\
  1 & \epsilon_{i}(\mathbf{k}) & \ldots & \epsilon_{i}^{i}(\mathbf{k}) & \ldots 
  & \epsilon_{i}^{n-1}(\mathbf{k})\\
  \vdots & \vdots & \ddots & \vdots & \ddots & \vdots\\
  1 & \epsilon_{n}(\mathbf{k}) & \ldots & \epsilon_{n}^{i}(\mathbf{k}) & \ldots
  & \epsilon_{n}^{n-1}(\mathbf{k})
 \end{array}
 \right)
 \left(
 \begin{array}{c}
  c_{0}(\mathbf{k})\\
  c_{1}(\mathbf{k})\\
  \vdots\\
  c_{i}(\mathbf{k})\\
  \vdots\\
  c_{n-1}(\mathbf{k})
  \end{array}
  \right)=-\left(
  \begin{array}{c}
   \epsilon_{1}^{n}(\mathbf{k})\\
   \vdots\\
   \epsilon_{i}^{n}(\mathbf{k})\\
   \vdots\\
   \epsilon_{n}^{n}(\mathbf{k})
  \end{array}
  \right) \, ,
  \label{eq:sislin}
 \end{equation}

where $\epsilon_{i}(\mathbf{k})$ represents the $i$-th energy band.

Therefore, using eigenvalues from the \textit{reference band structure}, 
we can solve this system to obtain the coefficients $c_i$ as 
functions of $\epsilon_i(\mathbf{k})$. This form of the coefficients will be 
called \textit{numerical functions}, denoted by the super-index $N$:
\begin{equation}
 \begin{cases}
  c^{N}_0\left[\epsilon_{1}(\mathbf{k}),\epsilon_{2}(\mathbf{k}),\ldots,\epsilon_{n}
  (\mathbf{k})\right]\\
  c^{N}_1\left[\epsilon_{1}(\mathbf{k}),\epsilon_{2}(\mathbf{k}),\ldots,\epsilon_{n}
  (\mathbf{k})\right]\\
  \vdots \\
  c^{N}_{n-1}\left[\epsilon_{1}(\mathbf{k}),\epsilon_{2}(\mathbf{k}),\ldots,\epsilon_{n}
  (\mathbf{k})\right] \, .
 \end{cases}
\end{equation}
Since we want to use the $\mathbf{k{\cdot}p}$ to describe our \textit{reference band 
structure}, we should now consider that both numerical and analytical forms of 
the coefficients are equivalent, leading to the equality
\begin{equation}
  \begin{cases}
   c^A_0(\mathbf{k},\{p\}) = c^{N}_0\left[\epsilon_{1}(\mathbf{k}),\epsilon_{2}
   (\mathbf{k}),\ldots,\epsilon_{n}(\mathbf{k})\right]\\
   c^A_1(\mathbf{k},\{p\}) = c^{N}_1\left[\epsilon_{1}(\mathbf{k}),\epsilon_{2}
   (\mathbf{k}),\ldots,\epsilon_{n}(\mathbf{k})\right]\\
   \vdots \\
   c^A_{n-1}(\mathbf{k},\{p\}) = c^{N}_{n-1}\left[\epsilon_{1}(\mathbf{k}),\epsilon_{2}
   (\mathbf{k}),\ldots,\epsilon_{n}(\mathbf{k})\right] \, .
  \end{cases}
  \label{eq:analyticalTOnumerical}
\end{equation}

Having both, analytical and numerical functions, we can perform the fitting 
procedure to extract the $\mathbf{k{\cdot}p}$ parameters that best describe the 
\textit{reference band structure}. 
The fitting was done using the nonlinear least squares method, 
implemented on Mathematica$^{\textsuperscript{TM}}$ using the NonLinearModelFit 
routine~\cite{Mathematica.NLLS}. Several different minimization methods were 
tested: Newton, QuasiNewton, LevenbergMarquardt, Gradient, Conjugate Gradient. 
As the results were similar for all tested methods, we chose the Conjugate 
Gradient method due to its relatively low memory requirements for a large-scale 
problem and simplicity of its iteration~\cite{adams1996linear}. 

The fitting method described above is general, and it can be applied for any 
given system, even for $\mathbf{k{\cdot}p}$ Hamiltonians larger than 
$8{\times}8$ and any direction in the FBZ.  For the particular case of 
semiconductors with zinc blende structures, we can sample the FBZ along the 
three most relevant high-symmetry directions, namely, ${\Gamma}-\text{X}$, 
${\Gamma}-\text{K}$, and ${\Gamma}-\text{L}$. The number of $\mathbf{k}$-point 
lines play an important role, e.g., the direction ${\Gamma}-\text{L}$ can yield 
only the $\tilde{\gamma}_1$ and $\tilde{\gamma}_3$ parameters, and hence, 
additional directions are required to identify the $\tilde{\gamma}_2$ 
parameter. 

For the case of the Hamiltonian given in eq. (\ref{eq:kpZB8}), the secular 
equation can always be factorized in the separate components, reducing the 
dimension of the problem by half. The factorized secular equation reads as 
\begin{equation}
 \begin{split}
 \left[ 
 c^N_0(\mathbf{k},\{p\}) 
 + c^N_1(\mathbf{k},\{p\})\epsilon 
 \right. & 
 + c^N_2(\mathbf{k},\{p\}) \epsilon^{2}
  \\
 & \left. 
 + c^N_3(\mathbf{k},\{p\})\epsilon^{3}
 + \epsilon^{4} 
 \right]^{2} = 0\, ,
 \end{split}
 \label{eq:factseceq}
\end{equation}
where $\{p\} = \{\tilde{\gamma}_1, \tilde{\gamma}_2, \tilde{\gamma}_3, \Delta_{so}, 
\text{P}, E_g, \tilde{\text{e}}\}$. 
In specific directions the secular equation may be further factorized. 

Solving the system (\ref{eq:factseceq}), we 
obtain the following relations for the numerical coefficients
\begin{equation}
  \begin{split}
   c^N_{HH} (k_{_{\Gamma X}}) = & \; - \epsilon_{HH}(k_{_{\Gamma X}}), \\
   c^N_2(k_{_{\Gamma X}}) = & \; -\epsilon_{CB}(k_{_{\Gamma X}})-\epsilon_{LH}(k_{_{\Gamma X}})-\epsilon_{SO}(k_{_{\Gamma X}}), \\
   c^N_1(k_{_{\Gamma X}}) = & \; \epsilon_{CB}(k_{_{\Gamma X}})\epsilon_{LH}(k_{_{\Gamma X}}) +\epsilon_{CB}(k_{_{\Gamma X}})\epsilon_{SO}(k_{_{\Gamma X}})\\+ &\epsilon_{LH}(k_{_{\Gamma X}})\epsilon_{SO}(k_{_{\Gamma X}}), \\
   c^N_0(k_{_{\Gamma X}}) = & \; \epsilon_{CB}(k_{_{\Gamma X}})\epsilon_{LH}(k_{_{\Gamma X}})
   \epsilon_{SO}(k_{_{\Gamma X}}).
  \end{split}
 \end{equation}
Notice that in the previous expressions, $\epsilon_1$, $\epsilon_2$, $\epsilon_3$ and $\epsilon_4$ where replaced by the average of the eigenvalues of the bands at the specific $k$-point: $\epsilon_{CB}$, $\epsilon_{HH}$, $\epsilon_{LH}$ and $\epsilon_{SO}$.

The parameters $\Delta_{so}$ and $E_g$ can be directly found from the 
$\Gamma$-point energies and used as input to the fitting approach. 
Since, we adjusted simultaneously the expressions for all the different bands 
in all chosen directions of the FBZ, the overall quality of the parameters for 
the multiband Hamiltonian is guaranteed.

\section{$\mathbf{k{\cdot}p}$ parameters for zinc blende G\MakeLowercase{a}A\MakeLowercase{s}} \label{sec:kpGaAs}

\begin{figure}[!htp]
\centering
\includegraphics[width=0.47\textwidth]{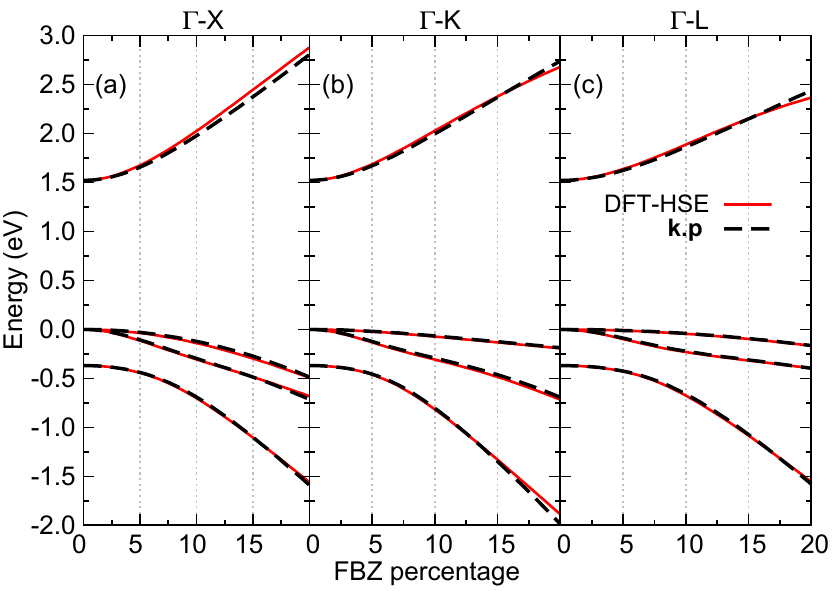}
\caption{Comparison between the band structure obtained by diagonalization of 
the $\mathbf{k{\cdot}p}$ Hamiltonian with the \SI{20}{\percent} region 
parameter set (dashed lines) and the band structure obtained by hybrid 
DFT-HSE+SOC (solid lines). We show three directions of the FBZ: (a) 
${\Gamma}-\text{X}$, (b) ${\Gamma}-\text{K}$ and (c) ${\Gamma}-\text{L}$. The 
x-axis shows percentage in the specific direction.} 
\label{fig:bs_8}
\end{figure}

In Fig.~\ref{fig:bs_8}, we show the results of the fitting, using 
\SI{20}{\percent} of the FBZ superposed to the original DFT-HSE+SOC 
calculation. For this particular range, we have found the following set of 
parameters: $\gamma_1 = 1.31$, $\gamma_2 = -0.72$, $\gamma_3 = 0.03$  and $e = 
-2.50$ in units of $\si{\planckbar^2/2 m_0}$; 
$\text{P} = 9.75~\si{\eV.\AA}$. A first inspection shows that the most 
important features of the band structure are preserved. The band structure for 
this range of wave vectors has essentially two different regions, one up to 
\SI{8}{\percent} of the FBZ and a second from \SIrange{8}{20}{\percent}. The HH 
and LH bands present nearly parabolic behavior in both regions, but the 
effective masses if calculated only inside each region, would be clearly 
different. The non-parabolicity, or band scattering,  around \SI{8}{\percent} 
and the quasi linear behavior of the conduction band and the split-off hole 
bands after the non-parabolicity are in good agreement with the {\it reference 
band structure}. Finally, a simple visual inspection of this results shows that 
the difference between the curves is smaller than \SI{8}{\percent} at the 
borders of the region. 

To avoid using visual estimation of the agreement of curves, it is necessary to 
find a procedure that numerically determines how close the DFT-HSE+SOC and the 
$8\times8$~$\mathbf{k{\cdot}p}$ band structures are with respect to wich other. 
This analysis can also be used to determine if in a smaller region, an 
optimized parameter can lead to more reliable results. To evaluate the 
agreement, we performed fittings over different ranges around the 
$\Gamma$-point, from \SI{2}{\percent} up to \SI{20}{\percent} 
of the FBZ, obtaining a large number of $\mathbf{k{\cdot}p}$ parameter sets. 

\section{Optimal parameter set} \label{sec:optimizationGaAs}

In order to evaluate the assertiveness of our parameters, we employed the Root 
Mean Square Deviation (RMSD) to compare the \textit{reference} 
and \textit{parametrized} band structures using the appropriate definition of the 
RMSD to our problem 
\begin{equation}
\text{RMSD}=\sqrt{\frac{1}{N}\sum_{d}^{N_d}\sum_{\mathbf{k}_d}^{N_\mathbf{k}}
\sum_{n}^{N_n} \left[\epsilon_{n}^{p}(\mathbf{k}_{d})-
\epsilon_{n}^{r}(\mathbf{k}_{d})\right]^{2}}
\label{eq:rmsd}
\end{equation}
where the summations run over the directions in which the FBZ was sampled, $d$, 
the points of the reciprocal space calculated in each direction, 
$\mathbf{k}_d$, and the bands taken into account, $n$. $N_d$, $N_\mathbf{k}$ 
and $N_n$ are the total values of each one of these variables. The super-index 
$p$ ($r$) in the energy bands denotes the \textit{parametrized} 
(\textit{reference}) band structure. Notice that the normalization condition 
(with $N = N_d \times N_\mathbf{k} \times N_n$) allows us to compare sets with 
different numbers of points. The smaller the value of the RMSD, the better our 
Hamiltonian and parameters fit the DFT-HSE+SOC band structure.

The search for the optimal parameter set is performed as follows: $(i)$ we 
determine the parameter sets for different fitting percentages of the FBZ; 
$(ii)$ for each of these parameter sets, we calculate the RMSD for different 
FBZ percentages; $(iii)$ the optimal parameter set presents the minimum RMSD 
value for a given FBZ percentage. We considered 16 different percentage values 
in the range from \SI{2}{\percent} to \SI{20}{\percent}, that were used to 
define either the parameter sets and the analyzed region. 

\begin{figure} 
\centering
\includegraphics[width=0.47\textwidth]{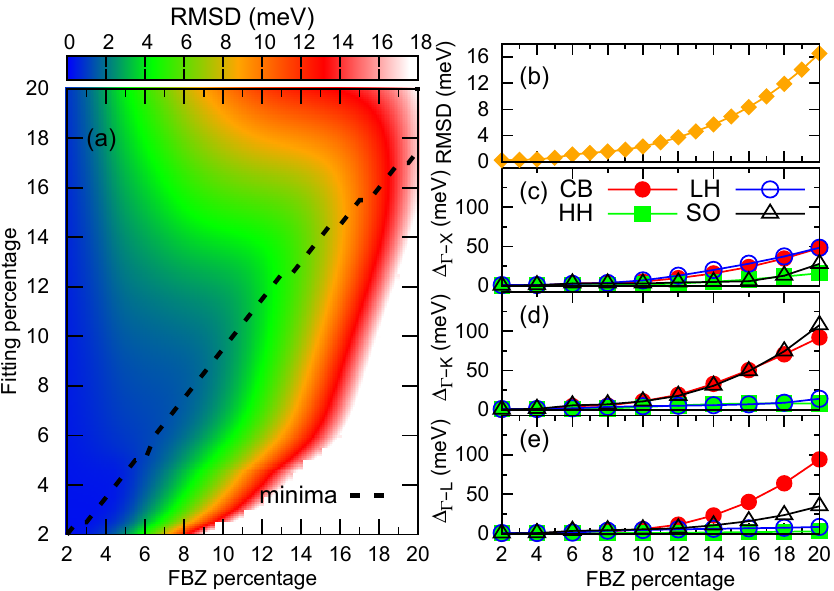}
\caption{(a) Root mean square deviation (RMSD) density map showing the 
agreement of the different adjusted parameter sets against the range around the 
$\Gamma$-point they are sampled. The optimal parameter sets are indicated by 
the dashed line. (b) RMSD of the optimal set 
of parameters for each enclosing region. (c), (d) and (e) show the maximum deviation 
for each optimal parameter set for the three directions used in the fitting process.}
\label{fig:rmsd}
\end{figure}

In Fig.~\ref{fig:rmsd}(a) we show the RMSD density map, with $y$-axis 
representing the fitting percentage of the parameter sets and the $x$-axis, the 
FBZ percentage used in the RMSD determination. The lowest RMSD values for each 
range are represented by the black dashed line. These parameter sets represent 
the best parameters that describe each range. We found that all parameter sets 
reproduce the band structure in the region below \SI{6}{\percent} with an 
average deviation of around \SI{2}{\meV}. If the parameter set is in the 
fitting range between \SIlist{6;14}{\percent}, the region of optimal agreement 
is extended to approximately \SI{12}{\percent} of the FBZ with just a slight 
increase of the RMSD value. If one considers a higher deviation, e.g. 
\SI{10}{\meV}, this region would be extended to around \SI{15}{\percent}. 
Animations of the optimal $\mathbf{k{\cdot}p}$ band structure changes with the 
fitting region limit can be found in the Supplemental Materials. 

Fig.~\ref{fig:rmsd}(b) shows the RMSD for the optimal parameters sets. We can 
see an increase of the average deviation by the increase of the FBZ range. This 
would be expected since the $8\times$8 $\mathbf{k{\cdot}p}$ Hamiltonian is 
valid in a region around $\Gamma$-point. The results presented here show that 
the average deviation for the \SI{20}{\percent} range is still below 
\SI{20}{\meV}, reasonable for most of the optical simulations and for ranges 
below \SI{14}{\percent} the average deviation is only \SI{4}{\meV}. 

The maximum deviation from the DFT-HSE calculation for each range in the 
\num{3} different directions, ${\Gamma}-\text{X}$, ${\Gamma}-\text{K}$ and 
${\Gamma}-\text{L}$, is shown in Figs.~\ref{fig:rmsd}(c)-(e), respectively. 
Although CB and SO present large deviations at \SI{20}{\percent} of the FBZ 
(approx. \SI{100}{\meV} along ${\Gamma}-\text{K}$ for CB and SO and also along 
${\Gamma}-\text{L}$ for CB), for all other sampled curves, the bands present up 
to this percentage a deviation much smaller (around \SI{50}{\meV} for CB and LH 
at ${\Gamma}-\text{X}$ and smaller than \SI{25}{\meV} for all others). The 
large values of the deviation for CB and SO, indicate that they are mainly 
responsible for the steep increase of the RMSD around \SI{15}{\percent}, i.e., 
all other curves have a very small deviation up to this percentage.
 
\begin{figure}[h]
\centering
\includegraphics[width=0.47\textwidth]{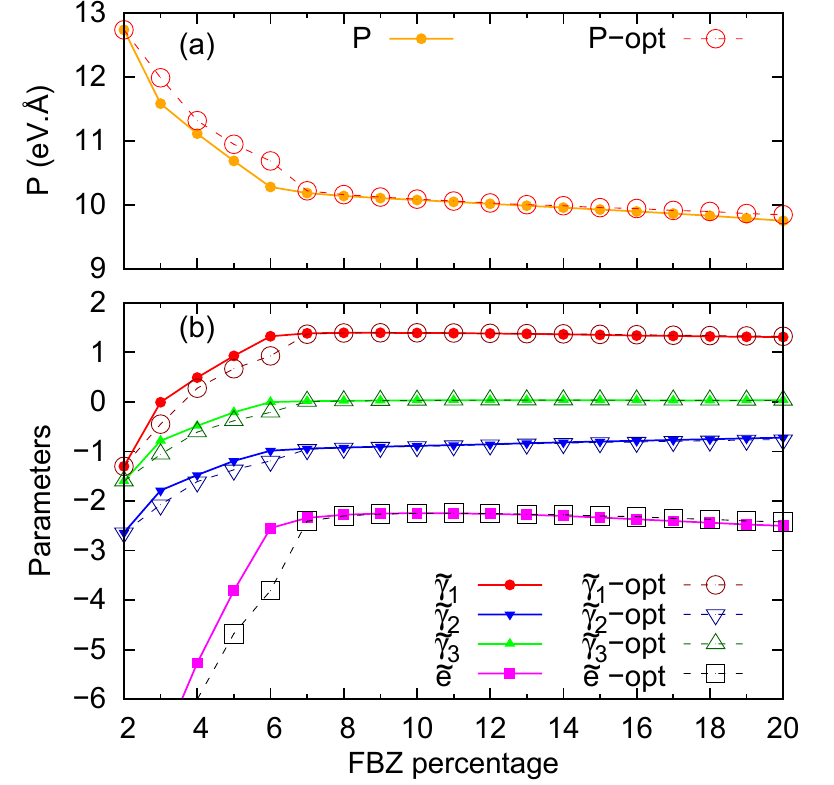}
\caption{ 
Comparison between optimal and non optimal parameter sets. (a) P parameter in 
$\si{\electronvolt.\angstrom}$ units and (b) $\tilde{\gamma}_1$, $\tilde{\gamma}_2$, 
$\tilde{\gamma}_3$ and $\tilde{e}$.}
\label{fig:par_8}
\end{figure}

A general overview of the parameter sets with and without the optimization 
approach is presented in Fig.~\ref{fig:par_8} for different FBZ regions. The 
dashed lines represent the raw data, i.e., the parameter sets obtained directly 
from the fitting of the specific range while the solid lines are used for the 
optimal parameters for the same range. One can clearly distinguish two 
different regions: i) below \SI{7}{\percent}, we can see a fast decay of the 
values for the interband interaction parameter, $P$, (on top) and a fast 
increase for the effectives masses (on bottom); and ii) above \SI{7}{\percent}, 
the parameters are almost stable with  a very slight linear variation.
 
\begin{figure}[h]
\centering
\includegraphics[width=0.47\textwidth]{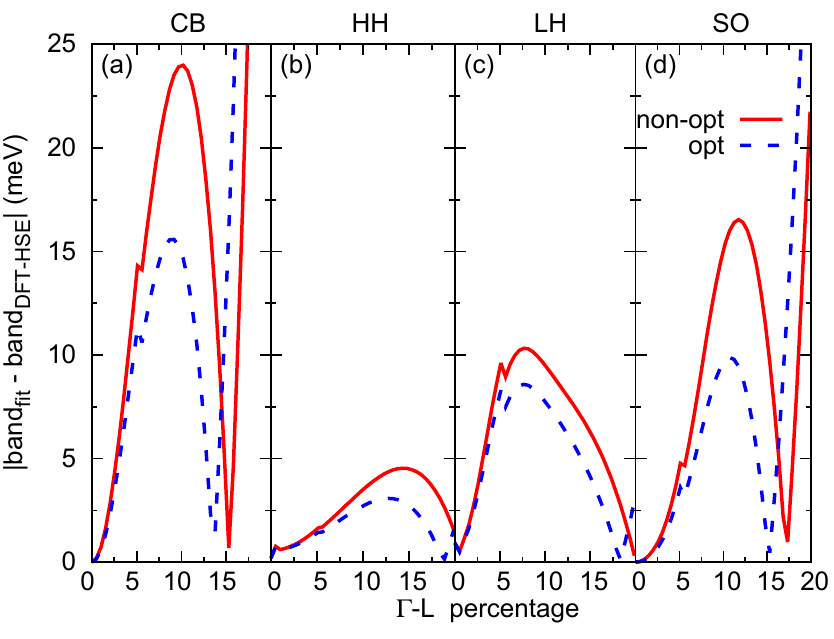}
\caption
{ 
 Difference between the DFT-HSE and $\mathbf{k{\cdot}p}$ band structures 
 calculated with optimal  (solid lines) and non-optimal (dashed lines) 
 parameter sets along $\Gamma-\text{L}$ direction for: (a) CB; (b) HH; (c) LH; 
 and (d) SO bands. The optimal  parameters set shows better agreement with the 
 DFT-HSE band structure. Other directions show similar behaviors.
 }
\label{fig:dif_bs}
 \end{figure}

Analyzing the band structure behavior, it is easy to notice that using a range 
that takes into account the non-parabolicity around \SI{8}{\percent} is 
essential to determine a stable set of parameters. In light of 
Fig.~\ref{fig:rmsd}(b) however, one can state that, even with the stability of 
the parameter values, an optimal set must be chosen to enhance the accuracy of 
the fitting. This can be seen on Fig.~\ref{fig:dif_bs}, where we present 
the agreement of \textit{parametrized} and \textit{reference} band structures 
for the optimal (solid lines) and non-optimal (dashed lines) parameter sets for 
the range of \SI{20}{\percent}. The optimal parameters for \SI{20}{\percent} 
were obtained for the fitting using the range of \SI{17.5}{\percent} and read 
as: $\tilde{\gamma}_1= 1.28 $, $\tilde{\gamma}_2 = -0.73$, $\tilde{\gamma}_3 = 
0.03$  and $\tilde{e} = -2.34$ and $\text{P} = 9.85 \si{\eV.\AA}$. 
Since the behavior of the bands in the different directions is very similar, we 
chose to present only the $\Gamma-\text{L}$ direction. The Supplemental 
Materials provide other directions expressions. The differences are more 
striking in conduction and split-off bands, where the choice of the parameters 
can reduce the total deviation to approximately two thirds for an specific 
point, i. e. from \SIrange[range-units=single]{25}{15}{\milli\electronvolt} on 
CB and from \SIrange[range-units=single]{15}{10}{\milli\electronvolt} on SO. 
To see a complete table with optimal parameters for the full range of enclosing 
regions, please refer to the Supplemental Materials.

\section{Comparison with literature parameters} \label{sec:comparison}
 
\begin{figure}[h]
\centering
\includegraphics[width=0.47\textwidth]{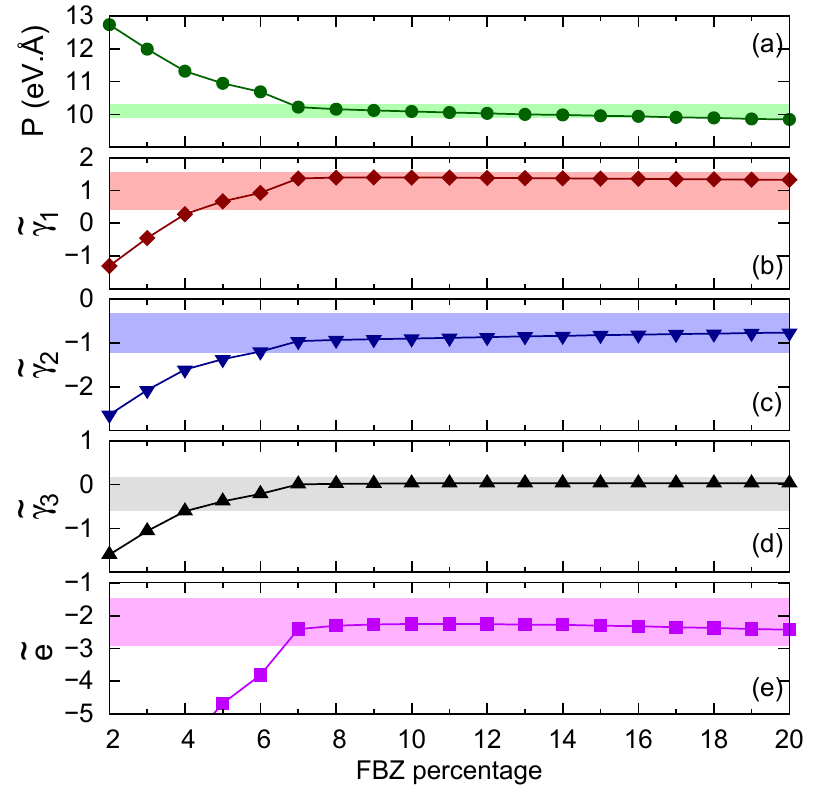}
\caption
{Comparison of the optimal parameters with the literature. The optimal 
parameters are shown in the curves and the shadowed regions present the 
intervals of the standard deviation around the average values of the 
parameters, determined from 7 traditional papers. (a) P parameter in 
$\si{\electronvolt.\angstrom}$, and (b) $\tilde{\gamma}_1$, 
(c) $\tilde{\gamma}_2$, (d) $\tilde{\gamma}_3$ and (e) $\tilde{e}$.
}
\label{fig:par_lit}
\end{figure}
The literature presents in general a unique set of parameters for any material. 
As we suggest optimal parameters for specific ranges of the FBZ, in our 
comparison we chose 7 different parameter sets from the 
literature~\cite{Vurgaftman2001,Shokhovets2007,Vrehen1968,Molenkamp1988,Lawaetz1971,Ostromek1996,Neumann1988}, 
see table in \ref{sec:appendixtable}. 
Using these sets, we calculated the average value for each parameter and its 
standard deviation. In Fig.~\ref{fig:par_lit} we plot the optimal parameters 
together with shadowed regions showing the intervals of the standard deviation 
around the average values of the parameters. Our results show good agreement 
with the literature data in general, since the values obtained for ranges 
larger than \SI{7}{\percent} are stable and lie always inside the standard 
deviation interval around the average of the values selected from the 
literature.

The behavior presented for regions smaller than \SI{7}{\percent} may be 
understood by a simple analysis the band structure and the role of $E_p$ in the 
secular equation. $E_P$ can be adimensionalized by defining a new parameter 
that reads as $\tilde{\gamma}_P = \nicefrac{E_P}{E_g}$, showing that, even if 
$P$ appears in first order perturbation terms, $\tilde{\gamma}_P$ acts as an 
effective mass parameter. According to this new definition, we have now five 
different effective mass parameters and four bands to do the fitting.
As up to \SI{7}{\percent}, the bands show a clear parabolic behavior, the 
fitting of the parameters become undetermined. Around this percentage all the 
bands start mixing and non-parabolic behavior may be seen. Just above this 
region, a new parabolic behavior emerges and all the bands change their 
curvatures accordingly. Including the two parabolic regions in the fitting, e. 
g., fitting from $\Gamma$ to \SI{12}{\percent}, provides the necessary 
relations to distinguish among the different parameters influence on the 
effective masses, giving parameters that agree with the literature parameters 
as can be seen in Fig.~\ref{fig:par_lit}. An evaluation of the method can be 
done by analysing the exceptional agreement with literature parameters.
The curvatures obtained by our fitting reproduce the most reliable data from 
literature. Moreover, this indicates that the choice of hybrid DFT-HSE combined 
methods reproduce accurately the properties of the actual electronic properties 
of the material, validating our choice.

Finally, joining the information of the agreement of the model with literature 
parameters together with the deviation from the DFT-HSE calculation described 
in section~\ref{sec:optimizationGaAs}, we have a tool to assess some insights 
about the accuracy of the effective mass approximation. The lack of agreement 
of the fitting after \SI{15}{\percent}, specially for the CB and SO bands, 
suggests that this specific approximation starts to lose its validity at this 
region. However, even in this region, our calculations indicate an average 
deviation of less than \SI{15}{\milli\eV}, indicating that, with proper 
parameters, the determination of properties depending on band structures inside 
this range of the FBZ are reliable. 

\section{Conclusions} \label{sec:conclusion}
 
We developed and implemented a general method to extract multiband 
$\mathbf{k{\cdot}p}$ parameters using the secular equation of the Hamiltonian. 
Our approach considers the simultaneous fitting of multiple directions of the 
FBZ of preexistent band structure and, combined with the RMSD  analysis, 
provides a tool to evaluate the global deviation between the fitted and the 
original data in a systematic way. Within this approach, an optimal set of 
parameters may be proposed for each specific region of the FBZ.

In order to test our approach, we fitted the conventional $8{\times}8$ zinc 
blende Hamitonian to GaAs band structure obtained by a state-of-the-art hybrid 
DFT-HSE+SOC calculation. The use of hybrid potentials provided a way of 
guaranteeing that the electronic properties of the systems are directly 
associated with their experimental values, addressing the most important 
issues when using DFT calculations to determine effective parameters. 

Our fitted band structures present good agreement with the DFT values when 
using up to \SI{20}{\percent} of FBZ. Particularly, below \SI{15}{\percent} we 
showed an average deviation of less than \SI{10}{\meV}. Above this range, we 
found that the deviation rapidly increases due to the lack of additional 
coupling terms in the Hamiltonian. Besides the good agreement on regions below 
\SI{7}{\percent}, our analysis show that the parameters are not stable in this 
range. The stability present above \SI{7}{\percent} and the small deviation 
below \SI{15}{\percent} define the range that can be used to obtain parameter 
sets that accurately describe the band structure up to \SI{20}{\percent} 
of the FBZ. Finally, the comparison with experimental and theoretical available 
data show that the optimal parameter sets lie inside the range of the most 
reliable parameters from the literature.

Concluding, our approach provides a method of finding parameters for a general 
$\mathbf{k{\cdot}p}$ model allowing its use for any phase or crystalline 
structure. As a consequence it can be used to extract parameters
of new $\mathbf{k{\cdot}p}$ Hamiltonians, opening a large range of 
opportunities to study new physical phenomena.

\section*{Acknowledgements}

The authors acknowledge financial support from the Brazilian agencies CNPq 
(grant \mbox{\#246549/2012-2}), FAPESP (grants~\#2011/19333-4, 
\#2012/05618-0 and \#2013/23393-8) and CAPES
(PVE grant \#88881.068174/2014-01).

 \appendix

 \section{Parameters table for comparison with the literature}
 \label{sec:appendixtable}
 In section \ref{sec:comparison} we compared  the optimal parameters
 for a region comprising up to \SI{20}{\percent} with the literature. 
 In table \ref{tab:par_8} we present
 the literature data used to calculate the average and standard deviation 
 of the parameters.
 
 \begin{table}
 	\protect\caption {
 	$\mathbf{k{\cdot}p}$ parameters obtained in this work (using \SI{20}{\percent} of the FBZ) and 
 	from selected references showing well established $\mathbf{k{\cdot}p}$ parameter sets. P is given in \si{\electronvolt \angstrom}.}
 \centering
  \begin{tabular}{cSSSSSSSS}
  \hline 
   & \multicolumn{1}{c|}{this work } & \multicolumn{7}{c}{literature}\tabularnewline
   \hline 
   & \multicolumn{1}{c}{fit (\SI{20}{\percent})} & \multicolumn{1}{c}{ref.\cite{Vurgaftman2001}}&
   \multicolumn{1}{c}{ref.\cite{Shokhovets2007}} & \multicolumn{1}{c}{ref. \cite{Lawaetz1971}}  &
   \multicolumn{1}{c}{ref. \cite{Ostromek1996}}  & \multicolumn{1}{c}{ref. \cite{Vrehen1968}}   & 
   \multicolumn{1}{c}{ref. \cite{Molenkamp1988}} & \multicolumn{1}{c}{ref. \cite{Neumann1988}} 
   \tabularnewline
    \hline 
    $\tilde{\gamma}_{1}$ &  1.28 &  0.66 &  0.36 &  2.02  & 0.60  &  0.98 &  0.47 &  1.21 \tabularnewline
    $\tilde{\gamma}_{2}$ & -0.73 & -1.10 & -1.08 & -0.41  & -1.17 & -0.61 & -1.24 & -0.10 \tabularnewline
    $\tilde{\gamma}_{3}$ &  0.03 &  0.23 & -0.45 &  0.46  & -0.37 & -0.61 & -0.48 & -0.07 \tabularnewline
    $\tilde{e}$          & -2.34 & -2.87 & -3.28 & -0.94  & -2.18 & -2.62 & -2.76 & -1.77 \tabularnewline
    $\text{P}$           &  9.85 & 10.47 & 10.25 &  9.89  & 10.27 & 10.37 & 10.48 & 10.18 \tabularnewline
    \hline 
   \end{tabular}
   \label{tab:par_8}
  \end{table}

\bibliographystyle{iopart-num} 
\bibliography{references}
\end{document}


\title{Supplementary Data to \\Stability and accuracy control of k.p 
parameters}

\author{Carlos M. O. Bastos}
\affiliation{S\~ao Carlos Institute of Physics, University of S\~ao Paulo, PO Box 369, 13560-970, S\~ao Carlos, SP, Brazil.}

\author{Fernando P. Sabino}
\affiliation{S\~ao Carlos Institute of Physics, University of S\~ao Paulo, PO Box 369, 13560-970, S\~ao Carlos, SP, Brazil.}

\author{Paulo E. Faria Junior}
\affiliation{S\~ao Carlos Institute of Physics, University of S\~ao Paulo, PO Box 369, 13560-970, S\~ao Carlos, SP, Brazil.}
\affiliation{Department of Physics, State University of New York at Buffalo, 14260, Buffalo, New York , USA}

\author{Tiago Campos}
\affiliation{S\~ao Carlos Institute of Physics, University of S\~ao Paulo, PO Box 369, 13560-970, S\~ao Carlos, SP, Brazil.}

\author{Juarez L. F. Da Silva}
\affiliation{S\~ao Carlos Institute of Chemistry, University of S\~ao Paulo, PO Box 780, 13560-970, S\~ao Carlos, SP, Brazil.}

\author{Guilherme M. Sipahi}
\affiliation{S\~ao Carlos Institute of Physics, University of S\~ao Paulo, PO Box 369, 13560-970, S\~ao Carlos, SP, Brazil.}
\affiliation{Department of Physics, State University of New York at Buffalo, 14260, Buffalo, New York , USA}

\maketitle

\section{The $\mathbf{k{\cdot}p}$ method} \label{sec:kp}

The quantum-mechanical treatment of the many body problem composed by electrons 
and nuclei in solid state materials is a complex task, in particular, due to 
the electron-electron interactions. Along the years, several approaches have 
been proposed to address this problem, which include the solution of the 
Schroedinger equation using trial wave functions at different levels of 
approximations such as the Hartree-Fock method combined with M{\o}ller-Plesset 
perturbation theory or methods based on DFT~\cite{Hohenberg1964,Kohn1965}. 
Although accurate, these methods are computationally demanding. The limit of 
thousand-atoms on one system using state-of-the-art computational resources 
precludes their use on mesoscopic and even in nanoscopic systems, e.g., 
a \SI{100}{\AA} wide zinc blende GaAs nanowire would demand at least \num{950} 
atoms for the correct description of one atomic layer. Plenty of interesting 
problems reside beyond this hard wall barrier. Alternatively, the use of the 
crystal symmetry to extract the main features of the electronic structure leads 
to another class of approaches that overcome the computational resources 
barrier, known generally as effective mass methods. In such methods, the 
many-body problem can be simplified by using an approximation in which an 
effective single electron moves in the field generated by the screened 
electron-nuclei and electron-electron systems. When many 
bands are included in this description the method is known as as the $\mathbf{k{\cdot}p}$ 
method, and have been used since the 50's 
\cite{Kane1966,Luttinger1955,Dresselhaus1955} to predict electronic and optical 
properties of semiconductors.

Below, we will summarize the key features of the $\mathbf{k{\cdot}p}$ method as 
it is described in several references elsewhere 
\cite{Enderlein1997,Willatzen2009,Kane1966}.
The one-electron Hamiltonian including relativistic SOC effects can be written 
as follows, 
\begin{equation}
H = \frac{p^2}{2m_0} + V(\mathbf{r}) + \frac{\hbar}{4m_{0}^{2}c^{2}}\left[
\mathbf{\nabla}V(\mathbf{r})\times\mathbf{p}\right] \cdot \boldsymbol{\sigma} \, ,
\label{eq:hamilPsi}
\end{equation}
where the first term is the kinetic energy of the electrons, the second term is 
the effective potential experienced by the electrons and the last term is the 
SOC contribution. The linear momentum operator is given by $\mathbf{p} = 
-i\hbar\nabla$, $m_0$ is the electron mass, $c$ is the velocity of light, 
$\hbar$ is the Planck constant divided by $2\pi$ and $\boldsymbol{\sigma}$ is a 
vector containing the Pauli matrices. Due to the translational symmetry of 
ideal crystalline systems, the effective potential is a periodic function, and 
hence, the wave function solution must satisfy the Bloch's theorem, i.e., 
\begin{equation}
\Psi_{n,\mathbf{k}}(\mathbf{r}) = e^{i\mathbf{k \cdot r}}u_{n,\mathbf{k}}(\mathbf{r}) \, ,
\label{eq:Psi}
\end{equation}
where $\Psi(\mathbf{r})$ is the total wave function (known as the Bloch function), 
$\mathbf{k}$ is a wave vector usually restricted to the FBZ, 
$u_{n,\mathbf{k}}(\mathbf{r})$ is a function with the same period as the 
crystal, and $n$ indicates the energy band index. A simple algebraic 
manipulation shows that 
\begin{equation}
\mathbf{p} \Psi_{n,\mathbf{k}}(\mathbf{r})
= e^{i\mathbf{k \cdot r}}(\hbar\mathbf{k}+\mathbf{p}) u_{n,\mathbf{k}}(\mathbf{r}) \, ,
\end{equation}
and by applying this transformation on the wave functions (\ref{eq:Psi}), the 
Hamiltonian  (\ref{eq:hamilPsi}), from now on identified as $H_{kp}$, may be 
simplified to act only on the periodic functions, 
$u_{n,\mathbf{k}}(\mathbf{r})$, i.e.,
\begin{equation}                                                               
H_{kp}\, u_{n,\mathbf{k}}(\mathbf{r}) = E_{n}(\mathbf{k})u_{n,\mathbf{k}}(\mathbf{r}) \, ,
\label{eq:H}                                                               
\end{equation}                                                                 
where,
\begin{equation}
 \begin{split}
 H_{kp} & = \frac{p^{2}}{2m_0} + V(\mathbf{r}) +
 \frac{\hbar^{2} k^2}{2m_0} + \frac{\hbar}{m_0}\mathbf{k} \cdot \mathbf{p} \\ 
 & + \frac{\hbar}{4m_{0}^{2}c^{2}}\left[\mathbf{\nabla}V(\mathbf{r})\times
 \mathbf{p}\right]\cdot\boldsymbol{\sigma} + \frac{\hbar^{2}}{4m_{0}^{2}c^{2}}
 \mathbf{k}\cdot\left[\boldsymbol{\sigma}\times\mathbf{\nabla}V(\mathbf{r})
 \right] \, .
\end{split}
\label{eq:H2}
\end{equation} 
Equation  (\ref{eq:H2}) is the $\mathbf{k{\cdot}p}$ Hamiltonian with 
SOC. This is an exact Hamiltonian that describes the motion of an electron in a 
periodic crystal. Despite been exact, there is no analytical solution for 
equation (\ref{eq:H}) and, at least, three approximations should be made in 
order to solve it to a certain degree.

The first approximation is to assume that we know the solutions for a 
particular point in reciprocal space, usually defined as $k_0$. Then, we can 
expand Hamiltonian (\ref{eq:H2}) around such point, and separate it into two 
different terms: one containing only non-vanishing terms at the expansion point
\begin{equation}
  H_0 = \frac{p^{2}}{2m_0} + V(\mathbf{r}) + \frac{\hbar^{2} k_0^2}{2m_0} + 
  \frac{\hbar}{m_0}\mathbf{k_0} \cdot \mathbf{p} + \frac{\hbar^{2}}{4m_{0}^{2}c^{2}}
  \mathbf{k_0}\cdot\left[\boldsymbol{\sigma}\times\mathbf{\nabla} V(\mathbf{r})\right] ,
\label{eq:H0}
\end{equation}
and the other containing the other terms
\begin{equation}
 \begin{split}
 H_P & = \frac{\hbar^{2}}{2m_0}\left(k^2 - k_0^2\right) + \frac{\hbar}{m_0} 
 \left(\mathbf{k}-\mathbf{k_0}\right) \cdot \mathbf{p} \\
 & + \frac{\hbar}{4m_{0}^{2}c^{2}}\left[\mathbf{\nabla}V(\mathbf{r})\times
 \mathbf{p}\right]\cdot\boldsymbol{\sigma} + \frac{\hbar^{2}}{4m_{0}^{2}c^{2}}
 \left(\mathbf{k}-\mathbf{k_0}\right)\cdot\left[\boldsymbol{\sigma}\times \mathbf{\nabla}V(\mathbf{r})\right] \, .
 \end{split}
\label{eq:HP}
\end{equation}
Rewriting equation (\ref{eq:H}), we get
\begin{equation}
H_{kp}\,u_{n,\mathbf{k}}(\mathbf{r}) =  \left(H_0 + H_P\right)u_{n,\mathbf{k}}(\mathbf{r}) = E_{n}(\mathbf{k}) u_{n,\mathbf{k}}(\mathbf{r})\, .
\label{eq:H0HP}
\end{equation}

The second approximation is to define a  basis set for equation 
(\ref{eq:H0HP}). In principle a complete basis set would be all orbitals on 
each atom of the basis of the crystal structure, i. e., any state from any atom 
of the crystal unit cell. Although this basis set is complete, it does not help 
on solving the problem, it is too big. An educated guess would be to 
use a truncated basis set that describe the most important features of the host 
crystal. Group theory is used to determine the symmetry of the states.

The third approximation is to use perturbation theory in order to define the 
matrix elements of equation (\ref{eq:H0HP}). A perturbative approach, proposed 
by L{\"o}wdin~\cite{Lowdin1951} in the early 50's is used to solve this problem.
In this formalism, the states are separated into two classes, A and B. The 
states in class A will be chosen in order to address the energy bands of 
interest and consequently, will be the basis set of the Hamiltonian matrix. 
Class B will comprise the remaining bands of the system. Even if the remote 
bands are outside the energy range we are interested in, their interaction with 
states in class A can provide important additional terms to the Hamiltonian. 
Using Dirac notation, from now on, a total state of the system can be written as
\begin{equation}
 |n\mathbf{k}\rangle = \sum_{\alpha}^A c_{\alpha n}(\mathbf{k})|\alpha\rangle+
 \sum_{\beta}^B c_{\beta n}(\mathbf{k})|\beta\rangle \, ,
 \label{eq:luw}
\end{equation}
where $|\alpha\rangle$ and $|\beta\rangle$ are the states in class A and B, 
respectively. For clarity, we have 
$u_{n,\mathbf{k}}(\mathbf{r}) = \langle n\mathbf{k}|\mathbf{r}\rangle$, 
$u_{\alpha,\mathbf{k_0}}(\mathbf{r}) = \langle\alpha|\mathbf{r}\rangle$ and 
$u_{\beta,\mathbf{k_0}}(\mathbf{r}) = \langle\beta|\mathbf{r}\rangle$. 
The symmetry provided in the previous step is used in this one to reduce the 
work by indicating the terms that are forbidden by symmetry.

Therefore, the matrix elements of equation (\ref{eq:H0HP}) are 
given by 
\begin{equation}
 \begin{split}
  \left\langle \alpha \left|H_0 + H_P\right| \alpha' \right\rangle &
  = E_{\alpha}\left(\mathbf{k-k_0}\right) \delta_{\alpha\alpha'}+ \langle \alpha | \frac{\hbar}{m_0}
  \left(\mathbf{k-k_0}\right)\cdot\mathbf{p} | \alpha' \rangle \\ 
  & + \langle \alpha | \frac{\hbar}{4m_{0}^{2}c^{2}}\left[\nabla V(\mathbf{r})
  \times\mathbf{p}\right]\cdot\boldsymbol{\sigma} | \alpha' \rangle  \\
  & +\sum_{\beta}\frac{\langle\alpha|\frac{\hbar}{m_0}\left(\mathbf{k-k_0}
  \right)\cdot\mathbf{p}|\beta\rangle \langle\beta|\frac{\hbar}{m_0}
  \left(\mathbf{k-k_0}\right)\cdot\mathbf{p}|\alpha'\rangle}{E_{\alpha}- E_{\beta}} \, ,
 \end{split}
\label{eq:matkp}
\end{equation}
with
\begin{equation}
 E_{\alpha}\left(\mathbf{k-k_0}\right) = E_\alpha(\mathbf{k_0})+\frac{\hbar^{2}}{2m_{0}} 
\left(k^{2}-k_0^2\right),
 \end{equation}
where $E_\alpha(\mathbf{k_0})$ is given by
\begin{equation}
 H_0u_{\alpha,\mathbf{k_0}}(\mathbf{r}) = E_\alpha(\mathbf{k_0})u_{\alpha,\mathbf{k_0}}(\mathbf{r}) \, .
\end{equation}

The evaluation of the matrix elements in equation (\ref{eq:matkp}) is indeed a 
very complicated task. Looking carefully into this expression, one can see that 
the dipole moments (proportional to the matrix elements  
$\langle\alpha|\frac{\hbar}{m_0}\left(\mathbf{k-k_0} \right)\cdot\mathbf{p}|\beta\rangle$)  
of all the transitions among the different states in the description are 
needed, as well as the transition energies associated with them ($E_\alpha$ and 
$E_\beta$). An alternative approach to look for all these data and performing 
all these sums, would be determining their functional form using group theory 
arguments~\cite{Kane1966,Dresselhaus1955,Dresselhaus2008}, 
replacing their analytical definitions by a parametrization. 

Our material of choice is the zinc blende GaAs that has a direct band gap with 
the maximum valence band (VBM) and the minimum conduction band (CBM) at the 
$\Gamma$-point. Thus, to investigate electronic properties such as optical 
transitions and transport, the choice of the $\Gamma$-point 
($\mathbf{k_0}=\mathbf{0}$) for the unperturbed Hamiltonian is straightforward. 
We considered as class A the following electronic states, the topmost six 
states in the valence band (VB) (usually referred to as $p$-like states) and 
the first two states at the conduction band (CB) (usually referred to as 
$s$-like states), 
as described below: 
\begin{equation}
\begin{aligned}[c]
  \left|\text{HH}\Uparrow\right\rangle   &
  =\frac{1}{\sqrt{2}}\left|\left(X+iY\right)\uparrow\right\rangle \\
  \left|\text{LH}\Uparrow\right\rangle  &
  =\frac{i}{\sqrt{6}}\left|\left(X+iY\right)\downarrow-2Z\uparrow\right\rangle \\
  \left|\text{SO}\Uparrow\right\rangle  &
  =\frac{1}{\sqrt{3}}\left|\left(X+iY\right)\downarrow+Z\uparrow\right\rangle \\
  \left|\text{CB}\Uparrow\right\rangle  & =
  \left|S\uparrow\right\rangle \\
 \end{aligned}
 \qquad \qquad
 \begin{aligned}[c]
 \left|\text{HH}\Downarrow\right\rangle  &
 =\frac{i}{\sqrt{2}}\left|\left(X-iY\right)\downarrow\right\rangle \\
 \left|\text{LH}\Downarrow\right\rangle  &
 =\frac{1}{\sqrt{6}}\left|\left(X-iY\right)\uparrow+2Z\downarrow\right\rangle \\
 \left|\text{SO}\Downarrow\right\rangle  &
 =\frac{i}{\sqrt{3}}\left|-\left(X-iY\right)\uparrow+Z\downarrow\right\rangle \\
 \left|\text{CB}\Downarrow\right\rangle  & =\left|S\downarrow\right\rangle \\
\end{aligned}\, ,
\label{eq:basekp}
\end{equation}
where HH, LH and SO are the heavy hole, light hole and split-off hole valence 
band states, respectively, and CB is the conduction band state. $\Uparrow$ and 
$\Downarrow$ represent a pseudo-spin variable used to distinguish the 
degenerate solutions at $\Gamma$-point. 

In the zinc blende symmetry group, $T_d$, the Bloch functions at $\Gamma$-point 
have the following symmetries~\cite{Dresselhaus2008,Yu2005a}: 
$\left| X \right\rangle \sim x$, $\left| Y \right\rangle \sim y$, 
$\left| Z \right\rangle \sim z$ and $\left| S \right\rangle \sim x^2 + y^2 + z^2$.  
The symbol $\sim$ means that the state on the left (e. g., $\left| X 
\right\rangle$) transforms as the function on the right (e. g., $x$-coordinate) 
under the symmetry operations of the $T_d$ group. The linear combinations of 
$\left| X \right\rangle$, $\left| Y \right\rangle$, $\left| Z \right\rangle$ and 
$\left| S \right\rangle$ given in the basis set (\ref{eq:basekp}) diagonalizes 
the SOC interaction at $\mathbf{k}=(0,0,0)$ \cite{Kane1957,Enderlein1997}. 
The matrix representation of (\ref{eq:matkp}) in the basis set 
(\ref{eq:basekp}) is
\begin{equation}
 \begin{pmatrix}
  Q & S & R & 0 & i\frac{S}{\sqrt{2}} & -i\sqrt{2}R & -iP_-& 0\\
  S^{\dagger} & T & 0 & R & i\frac{(T-Q)}{\sqrt{2}} & i\sqrt{\frac{3}{2}}S &
  \sqrt{\frac{2}{3}}P_z & -\frac{1}{\sqrt{3}}P_-\\
  R^{\dagger} & 0 & T & -S & -i\sqrt{\frac{3}{2}}S^{\dagger} & i\frac{(T-Q)}{\sqrt{2}} &
  -\frac{i}{\sqrt{3}}P_+ & -i\sqrt{\frac{2}{3}}P_z \\
  0 & R^{\dagger} & -S^{\dagger} & Q & -i\sqrt{2}R^{\dagger} & 
  -i\frac{S^{\dagger}}{\sqrt{2}}& 0 & -P_+\\
  -i\frac{S^{\dagger}}{\sqrt{2}} & -i\frac{(T-Q)^{\dagger}}{\sqrt{2}} &
  i\sqrt{\frac{3}{2}}S & i\sqrt{2}R & \frac{Q+T}{2}+\Delta_{so} & 0 & 
  -\frac{i}{\sqrt{3}}P_z &-i\sqrt{\frac{2}{3}}P_- \\
  i\sqrt{2}R^{\dagger} & -i\sqrt{\frac{3}{2}}S^{\dagger} &
  -i\frac{(T-Q)^{\dagger}}{\sqrt{2}} & i\frac{S}{\sqrt{2}} & 0 & 
  \frac{Q+T}{2}+\Delta_{so} & \sqrt{\frac{2}{3}}P_+ &  -\frac{1}{\sqrt{3}}P_z\\
  -iP_-& \sqrt{\frac{2}{3}}P_z  &  \frac{i}{\sqrt{3}}P_-& 0&  
  \frac{i}{\sqrt{3}}P_z & \sqrt{\frac{2}{3}}P_- & E_c & 0\\
  0 &  - \frac{1}{\sqrt{3}}P_+ & i\sqrt{\frac{2}{3}}P_z  & -P_- & 
  i\sqrt{\frac{2}{3}}P_+ &  -\frac{1}{\sqrt{3}}P_z & 0 & E_c \\ 
 \end{pmatrix}
\label{eq:kpZB8}
\end{equation}
where the terms are given by
\begin{equation}
 \begin{aligned}[l]
  Q&=-\frac{\hbar^2}{2 m_0}\left[(\tilde{\gamma}_{1}+\tilde{\gamma}_{2})(k_{x}^{2}+k_{y}^{2})-
  (\tilde{\gamma}_{1}-2\tilde{\gamma}_{2})\, k_{z}^{2}\right] &
  R&=-\frac{\hbar^2}{2 m_0}\sqrt{3}\left[
  \tilde{\gamma}_{2} (k_{x}^{2}-k_{y}^{2}) + 2 i \tilde{\gamma}_{3} k_{x} k_{y}
  \right]\\
  E_c&=E_g + \frac{\hbar^2}{2 m_0} \tilde{\text{e}}\, k^2 &
  P_z&=\text{P} \, k_z \\
  T&=-\frac{\hbar^2}{2 m_0}\left[(\tilde{\gamma}_{1}-\tilde{\gamma}_{2})(k_{x}^{2}+k_{y}^{2})+
  (\tilde{\gamma}_{1}+2\tilde{\gamma}_{2})\, k_{z}^{2} \right]&
  S&=i\frac{\hbar^2}{2 m_0}\left[2 \sqrt{3}\tilde{\gamma}_{3}k_{z}(k_{x}-ik_{y})\right]\\
  P_\pm&=\frac{1}{\sqrt{2}}\text{P}\left(k_x\pm ik_y\right) &
  k^2&=k_x^2+k_y^2+k_z^2\\
 \end{aligned}\ 
\end{equation}

 \section{Analytical functions for the $\Gamma$-$K$ and $\Gamma$-$L$ directions}
 \label{sec:appendix}
 

 In the main article, we presented the general procedure to determine analytical functions and their 
 numerical counterparts. In this appendix we show the functions for all directions used in this
 work.
 
 For the $\Gamma$-$X$ direction:
\begin{equation}
  \begin{split}
    c_{HH}^{A}(k_{_{_{\Gamma X}}},\{p\}) = & \;
    -(\tilde{\gamma}_{1}-2\tilde{\gamma}_{2})\frac{\hbar^{2} k_{_{\Gamma X}}^{2}}{2m_{0}}\\
    c_{0}^{A}(k_{_{\Gamma X}},\{p\}) \ \ \, = & \;
    \Delta_{_{SO}}-E_{g}
    +\left[ 
      2(\tilde{\gamma}_{1}+\tilde{\gamma}_{2})-\tilde{e}
    \right]
    \frac{\hbar^{2} k_{_{\Gamma X}}^{2}}{2m_{0}} \\
    c_{1}^{A}(k_{_{\Gamma X}},\{p\}) \ \ \, = & \;
    -\Delta_{_{SO}}E_{g}
    -\left[
      \Delta_{_{SO}}(\tilde{e}-\tilde{\gamma}_{1}-2\tilde{\gamma}_{2})
      +2E_{g}(\tilde{\gamma}_{1}+\tilde{\gamma}_{2})
      +\frac{2m_{0}}{\hbar^{2}}P^{2}
    \right]\frac{\hbar^{2} k_{_{\Gamma X}}^{2}}{2m_{0}}\\
    & \;
    + \left[
      -2\tilde{e}(\tilde{\gamma}_{1}+\tilde{\gamma}_{2})
      +\tilde{\gamma}_{1}^{2}+2\tilde{\gamma}_{1}\tilde{\gamma}_{2}-8\tilde{\gamma}_{2}^{2}
    \right]
    \left(\frac{\hbar^{2} k_{_{\Gamma X}}^{2}}{2m_{0}} \right)^{2}\\
    c_{2}^{A}(k_{_{\Gamma X}},\{p\}) \ \ \, = & \;
    -\left[
      \Delta_{_{SO}}E_{g}(\tilde{\gamma}_{1}+2\tilde{\gamma}_{2})+\frac{2}{3}\Delta_{_{SO}}\frac{2m_{0}}{\hbar^{2}}P^{2}
    \right]\frac{\hbar^{2}}{2m_{0}}k_{_{\Gamma X}}^{2}\\
    & \;
    - \left[
      \Delta_{_{SO}}\tilde{e}(\tilde{\gamma}_{1}+2\tilde{\gamma}_{2})
      +E_{g}(\tilde{\gamma}_{1}+4\tilde{\gamma}_{2})(\tilde{\gamma}_{1}-2\tilde{\gamma}_{2})
      +\frac{2m_{0}}{\hbar^{2}}P^{2}(\tilde{\gamma}_{1}-2\tilde{\gamma}_{2})
    \right]
    \left(\frac{\hbar^{2} k_{_{\Gamma X}}^{2}}{2m_{0}}\right)^{2}\\
    & \;
    - \tilde{e}(\tilde{\gamma}_{1}+4\tilde{\gamma}_{2})(\tilde{\gamma}_{1}-2\tilde{\gamma}_{2})
    \left(\frac{\hbar^{2} k_{_{\Gamma X}}^{2}}{2m_{0}}\right)^{3}\\
  \end{split}
 \end{equation}
where $k_{_{\Gamma X}}$ indicates a point along the $\Gamma - X$ direction. Please notice that in 
these expressions the Kane parameter $P$ appear always as part of its energetic counterpart 
$E_P = \nicefrac{2 m_0 P^2}{\hbar^2}$.

For the $\Gamma$-$K$ direction we have
 
 \begin{equation}
   \begin{split}
     c_{0}^{A}(k_{_{\Gamma K}},\{p\}) \ \ \, = & \;
     \Delta_{_{SO}}-E_{g}-(\tilde{e}-3\tilde{\gamma}_{1})\frac{\hbar^{2} k_{_{\Gamma K}}^{2}}{2m_{0}}\\
     c_{1}^{A}(k_{_{\Gamma K}},\{p\}) \ \ \, = & \;
     -\Delta_{_{SO}}E_{g}
     -\left[
       \Delta_{_{SO}}(\tilde{e}-2\tilde{\gamma}_{1})+3E_{g}\tilde{\gamma}_{1}+\frac{2m_{0}}{\hbar^{2}}P^{2}
     \right]
     \frac{\hbar^{2} k_{_{\Gamma K}}^{2}}{2m_{0}}\\
     & \;
     -\left[
       3\tilde{e}\tilde{\gamma}_{1}-3\tilde{\gamma}_{1}^{2}+3\tilde{\gamma}_{2}^{2}+9\tilde{\gamma}_{3}^{2}
     \right]
     \left(\frac{\hbar^{2} k_{_{\Gamma K}}^{2}}{2m_{0}}\right)^{2}\\
     c_{2}^{A}(k_{_{\Gamma K}},\{p\}) \ \ \, = & \;
     -\left[
       2\Delta_{_{SO}}E_{g}\tilde{\gamma}_{1}+\frac{2}{3}\Delta_{_{SO}}\frac{2m_{0}}{\hbar^{2}}P^{2}
     \right]
     \frac{\hbar^{2} k_{_{\Gamma K}}^{2}}{2m_{0}}\\
     & \;
     -\left[
       2\left(\tilde{e}\tilde{\gamma}_{1}-\tilde{\gamma}_{1}^{2}+\tilde{\gamma}_{2}^{2}+3\tilde{\gamma}_{3}^{2}\right) \Delta_{_{SO}}
       -3\left(-\tilde{\gamma}_{1}^{2}+\tilde{\gamma}_{2}^{2}+\tilde{\gamma}_{3}^{2}\right)E_{g}
       \right. \\
       & \;
       \ \ \ \ \left.
       + \left(2\tilde{\gamma}_{1}-\tilde{\gamma}_{2}-3\tilde{\gamma}_{3}\right)
       \frac{2m_{0}}{\hbar^{2}}P^{2}
     \right]
     \left(\frac{\hbar^{2} k_{_{\Gamma K}}^{2}}{2m_{0}}\right)^{2}\\
     & \;
     -\left[
       3\tilde{e} \left(\tilde{\gamma}_{1}^{2}-\tilde{\gamma}_{2}^{2}-3\tilde{\gamma}_{3}^{2}\right)
       +\tilde{\gamma}_{1}(-\tilde{\gamma}_{1}^{2}+3\tilde{\gamma}_{2}^{2}+9\tilde{\gamma}_{3}^{2})
       +2\tilde{\gamma}_{2}^{3}-18\tilde{\gamma}_{2}\tilde{\gamma}_{3}^{2}
     \right]
     \left(\frac{\hbar^{2} k_{_{\Gamma K}}^{2}}{2m_{0}}\right)^{3}\\
     c_{3}^{A}(k_{_{\Gamma K}},\{p\}) \ \ \, = & \;
     -\left[
       \Delta_{_{SO}}E_{g}\left(\tilde{\gamma}_{1}^{2}-\tilde{\gamma}_{2}^{2}-3\tilde{\gamma}_{3}^{2}\right)
       +\frac{\Delta_{_{SO}}}{3}(2\tilde{\gamma}_{1}-\tilde{\gamma}_{2}-3\tilde{\gamma}_{3})\frac{2m_{0}}{\hbar^{2}}P^{2}
     \right]
       \left(\frac{\hbar^{2} k_{_{\Gamma K}}^{2}}{2m_{0}}\right)^{2}\\
     & \;
     -\left[
       \Delta_{_{SO}}\tilde{e}\left(\tilde{\gamma}_{1}^{2}-\tilde{\gamma}_{2}^{2}-3\tilde{\gamma}_{3}^{2}\right)+E_{g}(\tilde{\gamma}_{1}-2\tilde{\gamma}_{2})(\tilde{\gamma}_{1}+\tilde{\gamma}_{2}+3\tilde{\gamma}_{3})(\tilde{\gamma}_{1}+\tilde{\gamma}_{2}-3\tilde{\gamma}_{3})
     \right. \\ 
     & \;
     \ \ \ \ +\left.
       (\tilde{\gamma}_{1}-2\tilde{\gamma}_{2})(\tilde{\gamma}_{1}+\tilde{\gamma}_{2}-3\tilde{\gamma}_{3})\frac{2m_{0}}{\hbar^{2}}P^{2}
     \right]
     \left(\frac{\hbar^{2} k_{_{\Gamma K}}^{2}}{2m_{0}}\right)^{3}\\
     & \;
     -\tilde{e}(\tilde{\gamma}_{1}-2\tilde{\gamma}_{2})(\tilde{\gamma}_{1}
     +\tilde{\gamma}_{2}+3\tilde{\gamma}_{3})(\tilde{\gamma}_{1}+\tilde{\gamma}_{2}-3\tilde{\gamma}_{3})
     \left(\frac{\hbar^{2} k_{_{\Gamma K}}^{2}}{2m_{0}}\right)^{4}
   \end{split}
 \end{equation}
 with the numerical counterpart being
 \begin{equation}
   \begin{split}
    c_{0}^{N}(k_{_{\Gamma K}})= & \; -\epsilon_{_{CB}}(k_{_{\Gamma K}})-\epsilon_{_{LH}}(k_{_{\Gamma K}})-\epsilon_{_{SO}}(k_{_{\Gamma K}})-\epsilon_{_{HH}}(k_{_{\Gamma K}})\\
    c_{1}^{N}(k_{_{\Gamma K}})= & \; \epsilon_{_{CB}}(k_{_{\Gamma K}})\epsilon_{_{LH}}(k_{_{\Gamma K}})+\epsilon_{_{SO}}(k_{_{\Gamma K}})\epsilon_{_{LH}}(k_{_{\Gamma K}})+\epsilon_{_{HH}}(k_{_{\Gamma K}})\epsilon_{_{LH}}(k_{_{\Gamma K}})+\epsilon_{_{CB}}(k_{_{\Gamma K}})\epsilon_{_{SO}}(k_{_{\Gamma K}})+\\
                                & \; \epsilon_{_{CB}}(k_{_{\Gamma K}})\epsilon_{_{HH}}(k_{_{\Gamma K}})+\epsilon_{_{SO}}(k_{_{\Gamma K}})\epsilon_{_{HH}}(k_{_{\Gamma K}})\\
    c_{2}^{N}(k_{_{\Gamma K}})= & \; -\epsilon_{_{CB}}(k_{_{\Gamma K}})\epsilon_{_{LH}}(k_{_{\Gamma K}})\epsilon_{_{SO}}(k_{_{\Gamma K}})-\epsilon_{_{CB}}(k_{_{\Gamma K}})\epsilon_{_{HH}}(k_{_{\Gamma K}})\epsilon_{_{SO}}(k_{_{\Gamma K}})-\epsilon_{_{LH}}(k_{_{\Gamma K}})\epsilon_{_{HH}}(k_{_{\Gamma K}})\epsilon_{_{SO}}(k_{_{\Gamma K}})-\\
                                & \; \epsilon_{_{CB}}(k_{_{\Gamma K}})\epsilon_{_{LH}}(k_{_{\Gamma K}})\epsilon_{_{HH}}(k_{_{\Gamma K}}) \\
    c_{3}^{N}(k_{_{\Gamma K}})= & \; \epsilon_{_{CB}}(k_{_{\Gamma K}})\epsilon_{_{LH}}(k_{_{\Gamma K}})\epsilon_{_{SO}}(k_{_{\Gamma K}})\epsilon_{_{HH}}(k_{_{\Gamma K}})
    \end{split}
 \end{equation}

The $\Gamma$-$L$ direction functions are 

\begin{equation}
  \begin{split}
    c_{HH}^{A}(k_{_{_{\Gamma L}}},\{p\}) = & \;
    -(\tilde{\gamma}_{1}-2\tilde{\gamma}_{3})\frac{\hbar^{2} k_{_{\Gamma L}}^{2}}{2m_{0}}\\
    c_{0}^{A}(k_{_{\Gamma L}},\{p\}) \ \ \, = & \;
    \Delta_{_{SO}}-E_{g}+
    \left[ 2(\tilde{\gamma}_{1}+\tilde{\gamma}_{3})-\tilde{e}\right]
    \frac{\hbar^{2} k_{_{\Gamma L}}^{2}}{2m_{0}}\\
    c_{1}^{A}(k_{_{\Gamma L}},\{p\}) \ \ \, = & \;
    -\Delta_{_{SO}}E_{g}
    -\left[
      \Delta_{_{SO}}(\tilde{e}-\tilde{\gamma}_{1}-2\tilde{\gamma}_{3})+2E_{g}(\tilde{\gamma}_{1}+\tilde{\gamma}_{3})+\frac{2m_{0}}{\hbar^{2}}P^{2}
    \right]
    \frac{\hbar^{2} k_{_{\Gamma L}}^{2}}{2m_{0}}\\
    & \;
    + \left[
    -2\tilde{e}(\tilde{\gamma}_{1}+\tilde{\gamma}_{3})
    +\tilde{\gamma}_{1}^{2}+2\tilde{\gamma}_{1}\tilde{\gamma}_{3}-8\tilde{\gamma}_{3}^{2}
    \right]
    \left(\frac{\hbar^{2} k_{_{\Gamma L}}^{2}}{2m_{0}}\right)^{2}\\
    c_{2}^{A}(k_{_{\Gamma L}},\{p\}) \ \ \, =& \;
    -\left[
      \Delta_{_{SO}}E_{g}(\tilde{\gamma}_{1}+2\tilde{\gamma}_{3})+\frac{2}{3}\Delta_{_{SO}}\frac{2m_{0}}{\hbar^{2}}P^{2}
    \right]
    \frac{\hbar^{2} k_{_{\Gamma L}}^{2}}{2m_{0}}\\
    & \;
    -\left[
    \Delta_{_{SO}}\tilde{e}(\tilde{\gamma}_{1}+2\tilde{\gamma}_{3})
    +E_{g}(\tilde{\gamma}_{1}+4\tilde{\gamma}_{3})(\tilde{\gamma}_{1}-2\tilde{\gamma}_{3})
    +\frac{2m_{0}}{\hbar^{2}}P^{2}(\tilde{\gamma}_{1}-2\tilde{\gamma}_{3})
    \right]
    \left(\frac{\hbar^{2} k_{_{\Gamma L}}^{2}}{2m_{0}}\right)^{2}\\
    & \;
    -\tilde{e}(\tilde{\gamma}_{1}+4\tilde{\gamma}_{3})(\tilde{\gamma}_{1}-2\tilde{\gamma}_{3})
    \left(\frac{\hbar^{2} k_{_{\Gamma L}}^{2}}{2m_{0}}\right)^{3}
  \end{split}
 \end{equation}
 
 and their numerical counterpart
 
 \begin{equation}
   \begin{split}
 c_{HH}^{N}(k_{_{\Gamma L}})=& \;-\epsilon_{_{HH}}(k_{_{\Gamma L}})\\
 c_{0}^{N}(k_{_{\Gamma L}})=& \;-\epsilon_{_{CB}}(k_{_{\Gamma L}})-\epsilon_{_{LH}}(k_{_{\Gamma L}})-\epsilon_{_{SO}}(k_{_{\Gamma L}})\\
 c_{1}^{N}(k_{_{\Gamma L}})=& \;\epsilon_{_{CB}}(k_{_{\Gamma L}})\epsilon_{_{LH}}(k_{_{\Gamma L}})+\epsilon_{_{SO}}(k_{_{\Gamma L}})\epsilon_{_{LH}}(k_{_{\Gamma L}})+\epsilon_{_{CB}}(k_{_{\Gamma L}})\epsilon_{_{SO}}(k_{_{\Gamma L}})\\
 c_{2}^{N}(k_{_{\Gamma L}})=& \;-\epsilon_{_{CB}}(k_{_{\Gamma L}})\epsilon_{_{LH}}(k_{_{\Gamma L}})\epsilon_{_{SO}}(k_{_{\Gamma L}})
   \end{split}
 \end{equation}

%

\section {6x6 Hamiltonian}

	In the paper, we considered the 8$\times$8 Kane Hamiltonian. However, for large gap materials or
	when the interest relies in effects occurring only inside the valence band, we can neglect the
	interaction between the conduction and valence bands, setting the parameter P to zero.
	In such approach, we can define two independent A classes, one for valence band (2$\times$2)
	states and the other for the conduction band states(6$\times$6), obtaining a new Hamiltonian
	that will be denoted as 6$\times$6~\cite{Willatzen2009}. 
	
	Although the functional form of the 6$\times$6 and 8$\times$8 terms are the same, the 
	different choices for the A classes requires correction in the effective mass parameters.
	The Hamiltonian is given by the following matrix
	\begin{equation}
		\begin{pmatrix}
			Q & S & R & 0 & i\frac{S}{\sqrt{2}} & -i\sqrt{2}R & 0& 0\\
			S^{\dagger} & T & 0 & R & i\frac{(T-Q)}{\sqrt{2}} & i\sqrt{\frac{3}{2}}S & 0 & 0\\
			R^{\dagger} & 0 & T & -S & -i\sqrt{\frac{3}{2}}S^{\dagger} & i\frac{(T-Q)}{\sqrt{2}} & 0 & 0 \\
			0 & R^{\dagger} & -S^{\dagger} & Q & -i\sqrt{2}R^{\dagger} & -i\frac{S^{\dagger}}{\sqrt{2}}& 0 & 0\\
			-i\frac{S^{\dagger}}{\sqrt{2}} & -i\frac{(T-Q)^{\dagger}}{\sqrt{2}} & i\sqrt{\frac{3}{2}}S & i\sqrt{2}R & \frac{Q+T}{2}+\Delta_{so} & 0 &  0 & 0\\
			i\sqrt{2}R^{\dagger} & -i\sqrt{\frac{3}{2}}S^{\dagger} & -i\frac{(T-Q)^{\dagger}}{\sqrt{2}} & i\frac{S}{\sqrt{2}} & 0 & \frac{Q+T}{2}+\Delta_{so} & 0 & 0\\
			0 & 0 &  0 & 0 &  0 & 0& E_c & 0\\
			0 & 0 & 0  &  0  & 0 &  0 & 0 & E_c \\ 
		\end{pmatrix}
		\label{eq:kpZB6}
	\end{equation}
	with the terms being
	\begin{equation}
		\begin{split}
		Q&=-\frac{\hbar^{2}} {2m_{0}}\left[(\gamma_{1}+\gamma_{2})(k_{x}^{2}+k_{y}^{2})-(\gamma_{1}-2\gamma_{2})k_{z}^{2}\right]\\
		R&=\frac{\hbar^{2}}{2m_{0}}\sqrt{3}\left[(2i\gamma_{3}k_{x}k_{y})+\gamma_{2}(k_{x}^{2}-k_{y}^{2})\right]\\
		E_c&=E_g + e\frac{\hbar^{2}k^2}{2m_{0}}
		\end{split}
		\quad\ \quad\
		\begin{split}
		T&=\frac{\hbar^{2}} {2m_{0}}\left[(\gamma_{2}-\gamma_{1})(k_{x}^{2}+k_{y}^{2})-(\gamma_{1}+2\gamma_{2})k_{z}^{2}\right]\\
		S&=\frac{\hbar^{2}} {2m_{0}}\left[2\sqrt{3}i\gamma_{3}k_{z}(k_{x}-ik_{y})\right]\\
		k^2&=k_x^2+k_y^2+k_z^2 \, .
		\end{split}
	\end{equation}

	\section{Fitting functions of the 6x6 Hamiltonian}

	  We also applied the fitting method to obtain the parameters for the 6$\times$6 Hamiltonian
	(\ref{eq:kpZB6}) as described in section 3 in the paper. In this Hamiltonian,
	valence and conduction band are decoupled and can be treated
	independently. As conduction band is a diagonal block with dimension 2, 
	we have the following polynomial equation:
	\begin{equation}
		\left[-\epsilon+c^A_0(\mathbf{k},\{p\})\right]^{2}=0 \, ,
	\end{equation}
	where the square in the expression means a the two-fold degeneracy of the eigenvalues.
	The analytical coefficient is then given by
	\begin{equation}
	   c^A_0(\mathbf{k},\{p\}) = E_g + e\frac{\hbar^{2}k^2}{2m_{0}},
		\label{eq:cA_CB_6}
	\end{equation}
	and the parameter set, $\{p\}$, in this case, is $\{e,E_g\}$.
	
	$e$ is determined through the fitting of the conduction band to a parabolic curve (\ref{eq:cA_CB_6}) and $E_g$ is extracted from the DFT-HSE band structure. For the valence band, also two-fold degenerated, we have the following secular equation
	\begin{equation}
	   \left[ \epsilon^{3}+ c^N_2(\mathbf{k},\{p\})\epsilon^{2}+c^N_1(\mathbf{k},\{p\})\epsilon + c^N_0(\mathbf{k},\{p\}) \right]^{2} = 0,
	\end{equation}
	with the parameter set being $\{p\}=\{\gamma_{1},\gamma_{2},\gamma_{3},\Delta_{so}\}$.

	The sampling directions were chosen to be the same used in the 8$\times$8 model, i. e., $\Gamma-X$, 
	$\Gamma-K$ and $\Gamma-L$. The analytical coefficients for the $\Gamma-X$ direction are
\begin{equation}
   \begin{split}
    c_{HH}^{A}(k_{_{_{\Gamma X}}},\{p\}) = & -\left[\gamma_{1}-2\gamma_{2}\right]\frac{\hbar^{2} k_{_{\Gamma X}}^{2}}{2m_{0}}\\
    c_{0}^{A}(k_{_{_{\Gamma X}}},\{p\}) = & \Delta_{so}+2\left[\gamma_1+\gamma_2\right]\frac{\hbar^{2} k_{_{\Gamma X}}^{2}}{2m_{0}}\\
    c_{1}^{A}(k_{_{_{\Gamma X}}},\{p\}) = & \Delta_{so}\left[\gamma_{1}+2\gamma_{2}\right]\frac{\hbar^{2} k_{_{\Gamma X}}^{2}}{2m_{0}}+\;
    \left[\gamma_1^2+2\gamma_{1}\gamma_{2} -8\gamma_{2}^2\right]\left(\frac{k_{_{\Gamma X}}^{2}}{2m_{0}}\right)^2
  \end{split}
   \label{eq:analyticalZB6_GX}
\end{equation}
and the numerical coefficients
\begin{equation}
  \begin{split}
   c^N_{HH} (k_{_{\Gamma X}}) = & \; \epsilon_{HH}(k_{_{\Gamma X}}) \\
   c^N_1(k_{_{\Gamma X}}) = & \;-\epsilon_{LH}(k_{_{\Gamma X}})-\epsilon_{SO}(k_{_{\Gamma X}})\\
   c^N_0(k_{_{\Gamma X}}) = & \;\epsilon_{LH}(k_{_{\Gamma X}}) \epsilon_{SO}(k_{_{\Gamma X}}) \ .
   \end{split}
   \end{equation}
For the $\Gamma-K$ direction we have
\begin{equation}
   \begin{split}
	   c_{0}^{A}(k_{_{_{\Gamma K}}},\{p\}) = & \Delta_{so}+3\gamma_{1}\frac{\hbar^{2} k_{_{\Gamma K}}^{2}}{2m_{0}}\\
	   c_{1}^{A}(k_{_{_{\Gamma K}}},\{p\}) = & 2\Delta_{so}\gamma_{1}\frac{\hbar^{2} k_{_{\Gamma K}}^{2}}{2m_{0}}+\;
	   3\left[\gamma_{1}^2-\gamma_{2}^2-3\gamma_{3}^2\right]\left(\frac{\hbar^{2} k_{_{\Gamma K}}^{2}}{2m_{0}}\right)^2\\
	   c_{2}^{A}(k_{_{_{\Gamma K}}},\{p\}) = & \Delta_{so}\left[\gamma_{1}^2-\gamma_{2}^2-3\gamma_{3}^2\right]\left(\frac{\hbar^{2} k_{_{\Gamma K}}^{2}}{2m_{0}}\right)^2+\;
	   \left(\gamma_{1}-2\gamma_{2}\right)\left[\gamma_{1}^2+2\gamma_{1}\gamma_{2}+\gamma_{2}^2-9\gamma_{3}^2\right]\left(\frac{\hbar^{2} k_{_{\Gamma K}}^{2}}{2m_{0}}\right)^3\
  \end{split}
   \label{eq:analyticalZB6_GK}
\end{equation}
with the numerical coefficients
\begin{equation}
  \begin{split}
    c^N_2(k_{_{\Gamma K}}) = & \; -\epsilon_{HH}(k_{_{\Gamma K}})-\epsilon_{LH}(k_{_{\Gamma K}})-\epsilon_{SO}(k_{_{\Gamma K}}) \\
   c^N_1(k_{_{\Gamma K}}) = & \; \epsilon_{HH}(k_{_{\Gamma K}})\epsilon_{LH}(k_{_{\Gamma K}}) +\epsilon_{HH}(k_{_{\Gamma K}})\epsilon_{SO}(k_{_{\Gamma K}})+ \epsilon_{LH}(k_{_{\Gamma K}})\epsilon_{SO}(k_{_{\Gamma K}}) \\
   c^N_0(k_{_{\Gamma K}}) = & \; \epsilon_{HH}(k_{_{\Gamma K}})\epsilon_{LH}(k_{_{\Gamma K}} \epsilon_{SO}(k_{_{\Gamma K}})\ .
   \end{split}
    \end{equation}
	And for the $\Gamma-L$ direction we have 
\begin{equation}
   \begin{split}
    c_{HH}^{A}(k_{_{_{\Gamma L}}},\{p\}) = & -\left[\gamma_{1}-2\gamma_{3}\right]\frac{\hbar^{2} k_{_{\Gamma L}}^{2}}{2m_{0}}\\
    c_{0}^{A}(k_{_{_{\Gamma L}}},\{p\}) = & \Delta_{so}+2\left[\gamma_1+\gamma_3\right]\frac{\hbar^{2} k_{_{\Gamma L}}^{2}}{2m_{0}}\\
    c_{1}^{A}(k_{_{_{\Gamma L}}},\{p\}) = & \Delta_{so}\left[\gamma_{1}+2\gamma_{3}\right]\frac{\hbar^{2} k_{_{\Gamma L}}^{2}}{2m_{0}}+\;
    \left[\gamma_1^2+2\gamma_{1}\gamma_{3} -8\gamma_{3}^2\right]\left(\frac{k_{_{\Gamma L}}^{2}}{2m_{0}}\right)^2
  \end{split}
   \label{eq:analyticalZB6_GL}
\end{equation}
with the numerical coefficients
\begin{equation}
  \begin{split}
   c^N_{HH} (k_{_{\Gamma L}}) = & \; \epsilon_{HH}(k_{_{\Gamma L}}) \\
   c^N_1(k_{_{\Gamma L}}) = & \;-\epsilon_{LH}(k_{_{\Gamma L}})-\epsilon_{SO}(k_{_{\Gamma L}}) \\
   c^N_0(k_{_{\Gamma L}}) = & \;\epsilon_{LH}(k_{_{\Gamma L}}) \epsilon_{SO}(k_{_{\Gamma L}})\ .
   \end{split}
   \end{equation}

\section{Parameter sets for the 6x6 Hamiltonian}
	
\begin{figure}[!htbp]
		\centering
		\includegraphics[width=0.9 \textwidth]{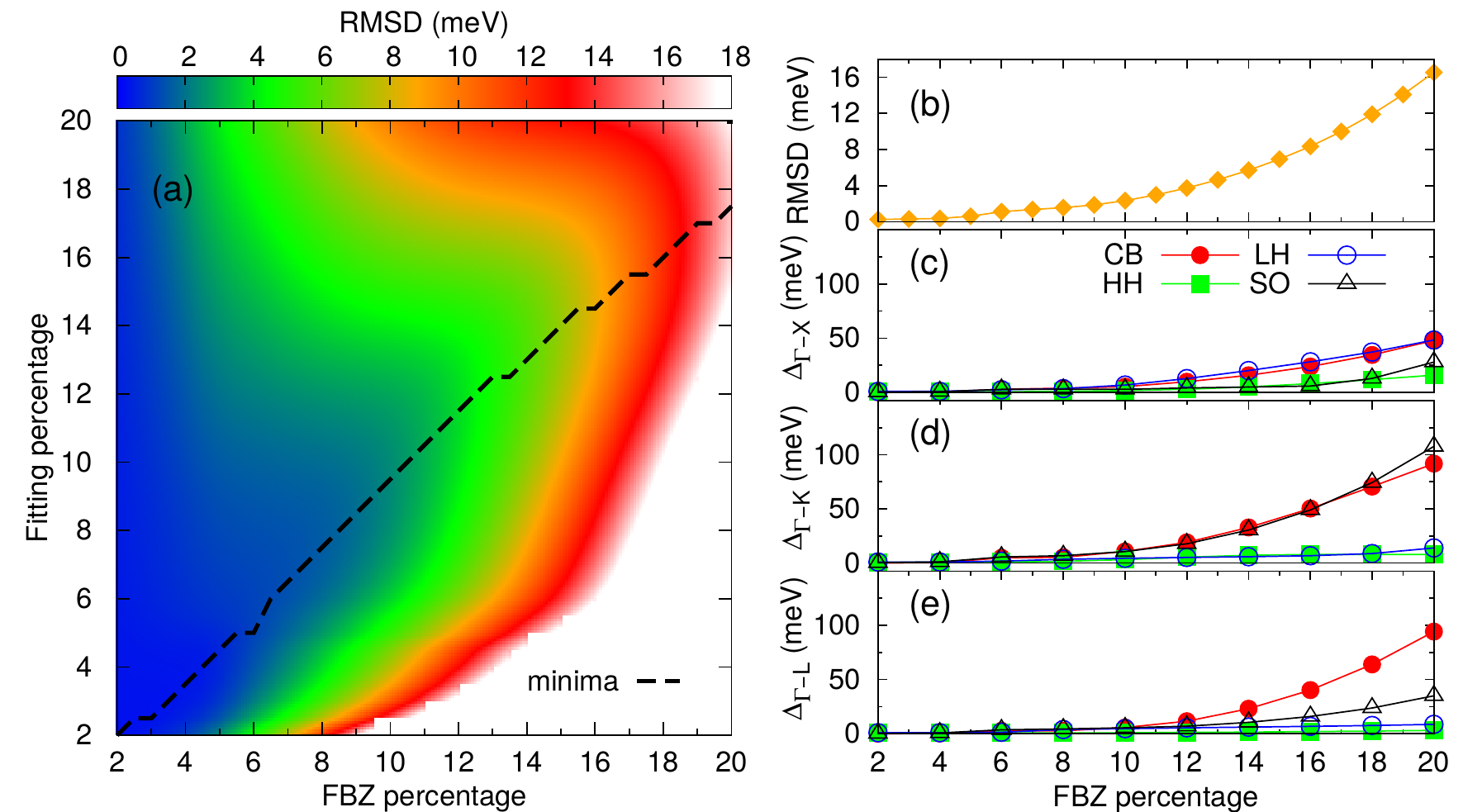}
		\label{fig:map_GaAs6}
		\caption{(a) Root mean square deviation (RMSD) values intensity map showing the agreement of the different adjusted parameter sets against the range around the $\Gamma$-point.
			A dashed curve indicate the optimal parameters 
			for each enclosed region. 
			(b) RMSD of the optimal set of parameters for each region.
			(c), (d) and (e) show the maximum deviation for each optimal parameter set
			for the three directions.}
	\end{figure}

	Using the same procedure as for the 8$\times$8 case, we performed the fitting for different
	regions around the $\Gamma$ point obtaining a different set of parameters for each one of them. 
	The definition of the best set of parameters was done using the RMSD analysis. 
	Fig. \ref{fig:map_GaAs6} shows the values of RMSD for each different region enclosing 
	the $\Gamma$ point. 

   	As expected, the lower RMSD values were obtained for fitting regions below \SI{8}{\percent}, where the band structure is almost parabolic. However, even for fitting regions beyond this limit, up to \SI{12}{\percent} of the FBZ, the RMSD values are still considerably small, indicating that the model can still be used.

	An animated figure showing the comparison of the $\mathbf{k{\cdot}p}$ fitting  and the original DFT-HSE 
	band structures is available with this text at http:// magazine site. In this animation we vary the 
	parameters showing emphasizing the fitting region used for their extraction.
	
	\newpage
		\begin{figure}[!htbp]
			\centering
			\includegraphics[width=0.7 \textwidth]{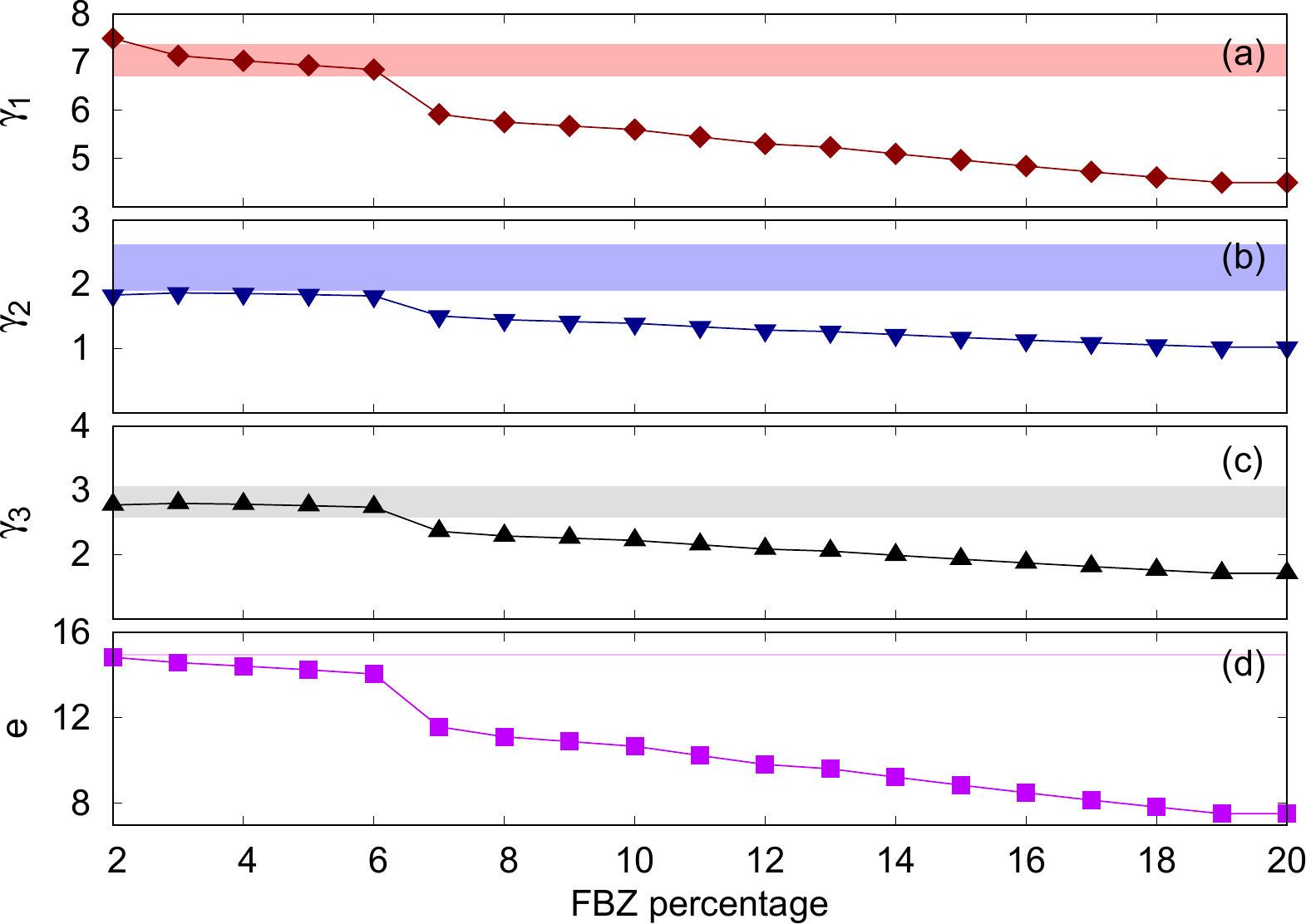}
			\caption{Comparison of the optimal parameters with the literature. The optimal    parameters 	are shown in the curves and the shadowed regions
				 present the intervals of the standard 
				deviation around the average values of the parameters, determined from 7 traditional papers 
				~\cite{Vurgaftman2001,Shokhovets2007,Lawaetz1971,Ostromek1996,Vrehen1968,Molenkamp1988,Neumann1988} 
				(see Table~\ref{tab:litpar_6}).
				 (a) $\gamma_1$, 
				(b) $\gamma_2$,
				 (c) $\gamma_3$
				  and (d) $e$.}
			\label{fig:par_6}
		\end{figure}

	Figure \ref{fig:par_6} shows the comparison of the optimal parameters with literature values.
	The colored regions indicate the region of one standard deviation around the average calculated with Refs.
	[\onlinecite{Vurgaftman2001,Shokhovets2007,Lawaetz1971,Ostromek1996,Vrehen1968,Molenkamp1988,Neumann1988}].
	For small percentage of the band (up to 8\%), the variation of the parameters is flat and similar 
	the literature, i.e., for the unique set parameters we describing this region.
	 However, above to 8\% (outside of the validity 
	of the Hamiltonian), the parameters have variation because the developed method return the best 
	values to describe effects non-present in the Hamiltonian. 

\begin{table}[!htbp]
\protect\caption {Experimental data used to compute the average and standard deviation of literature parameters}
\centering

\begin{tabular}{cSSSSSSSS}
	\hline \hline
& \multicolumn{1}{c}{ref.[\onlinecite{Vurgaftman2001}]} & \multicolumn{1}{c}{ref.[\onlinecite{Shokhovets2007}]}
& \multicolumn{1}{c}{ref. [\onlinecite{Lawaetz1971}]} & \multicolumn{1}{c}{ref. [\onlinecite{Ostromek1996}]}
& \multicolumn{1}{c}{ref.[\onlinecite{Vrehen1968}]} & \multicolumn{1}{c}{ref. [\onlinecite{Molenkamp1988}]}
& \multicolumn{1}{c}{ref. [\onlinecite{Neumann1988}]} \tabularnewline
\hline 
$\gamma_{1}$ &  6.98 &  6.85 &  7.65 & 6.67   & 7.20   &  6.79  & 7.17  \tabularnewline
$\gamma_{2}$ &  2.06 &  2.16 &  2.41 & 1.87   & 2.50   &  1.92  & 2.88  \tabularnewline
$\gamma_{3}$ &  2.93 &  2.79 &  3.28 & 2.67   & 2.50   &  2.68  & 2.91  \tabularnewline
$e$          & 14.93 & 14.93 & 14.93 & 14.93  & 14.93  & 15.04  & 15.04 \tabularnewline
\hline
\end{tabular}
\label{tab:litpar_6}
\end{table}

\section {Parameters for regions defined by different percentages of the FBZ}

\begin{table}[!htbp]
	\centering
\caption{Optimal parameter sets for different enclosing regions of the FBZ, for both Hamiltonians (6$\times$6 and 8$\times$8).}
\begin{tabular}{cSSSS SSSSSSS}

	\hline \hline
	&& \multicolumn{4}{c}{6$\times$6  Hamiltonian} && \multicolumn{5}{c}{8$\times$8  Hamiltonian}\tabularnewline
	\cline{3-6}\cline{8-12}
	\% &&\multicolumn{1}{c}{$\gamma_{1}$} & \multicolumn{1}{c}{$\gamma_{2}$} &\multicolumn{1}{c}{ $\gamma_{3}$}
	&\multicolumn{1}{c}{$\text{e}$} && \multicolumn{1}{c}{$\tilde{\gamma}_{1}$} &\multicolumn{1}{c}{ $\tilde{\gamma}_{2}$}
	&\multicolumn{1}{c}{ $\tilde{\gamma}_{3}$} &\multicolumn{1}{c}{ $\tilde{\text{e}}$} & \multicolumn{1}{c}{$\text{P}$}\tabularnewline
	\hline 
	2  && 7.49 & 1.83 & 2.77 & 14.82 &&-1.26 & -2.55 & -1.55 &-11.29 & 12.74  \tabularnewline
	3  && 7.13 & 1.86 & 2.80 & 14.57 &&-0.43 & -2.00 & -1.02 & -8.30 & 11.99  \tabularnewline
	4  && 7.03 & 1.85 & 2.78 & 14.41 && 0.27 & -1.55 & -0.59 & -5.80 & 11.32  \tabularnewline
	5  && 6.93 & 1.84 & 2.76 & 14.24 && 0.64 & -1.32 & -0.37 & -4.52 & 10.95  \tabularnewline
	6  && 6.85 & 1.82 & 2.74 & 14.04 && 0.89 & -1.15 & -0.21 & -3.68 & 10.69  \tabularnewline
	7  && 5.92 & 1.50 & 2.36 & 11.57 && 1.32 & -0.92 &  0.01 & -2.33 & 10.22  \tabularnewline
	8  && 5.76 & 1.45 & 2.29 & 11.12 && 1.34 & -0.90 &  0.02 & -2.22 & 10.16  \tabularnewline
	9  && 5.67 & 1.42 & 2.26 & 10.89 && 1.35 & -0.88 &  0.02 & -2.19 & 10.12  \tabularnewline
	10 && 5.60 & 1.39 & 2.22 & 10.67 && 1.35 & -0.87 &  0.03 & -2.17 & 10.09  \tabularnewline
	11 && 5.45 & 1.34 & 2.15 & 10.24 && 1.35 & -0.85 &  0.03 & -2.17 & 10.06  \tabularnewline
	12 && 5.30 & 1.29 & 2.09 &  9.82 && 1.34 & -0.83 &  0.03 & -2.18 & 10.03  \tabularnewline
	13 && 5.23 & 1.26 & 2.05 &  9.62 && 1.33 & -0.82 &  0.03 & -2.19 & 10.00  \tabularnewline
	14 && 5.10 & 1.22 & 1.99 &  9.23 && 1.32 & -0.81 &  0.03 & -2.20 &  9.99  \tabularnewline
	15 && 4.97 & 1.17 & 1.93 &  8.86 && 1.31 & -0.79 &  0.03 & -2.22 &  9.96  \tabularnewline
	16 && 4.84 & 1.13 & 1.87 &  8.50 && 1.31 & -0.78 &  0.03 & -2.24 &  9.94  \tabularnewline
	17 && 4.73 & 1.09 & 1.81 &  8.16 && 1.21 & -0.77 &  0.03 & -2.27 &  9.91  \tabularnewline
	18 && 4.61 & 1.05 & 1.76 &  7.83 && 1.29 & -0.76 &  0.03 & -2.29 &  9.90  \tabularnewline
	19 && 4.50 & 1.02 & 1.71 &  7.52 && 1.28 & -0.74 &  0.03 & -2.32 &  9.86  \tabularnewline
	20 && 4.50 & 1.02 & 1.71 &  7.52 && 1.28 & -0.73 &  0.03 & -2.34 &  9.85  \tabularnewline
	\hline 
\label{tab:allpar}
\end{tabular}
\end{table}

\bibliographystyle{elsarticle-num} 
\bibliography{references}